\documentclass[11pt]{article}

\usepackage{arxiv}

\usepackage[utf8]{inputenc} % allow utf-8 input
\usepackage[T1]{fontenc}    % use 8-bit T1 fonts
\usepackage{url}            % simple URL typesetting
\usepackage{booktabs}       % professional-quality tables
\usepackage{amsthm,amsmath,amsfonts}
\usepackage{natbib}
\usepackage{float}
\usepackage{subcaption}
\usepackage{algorithm2e}
\usepackage{xcolor}
\usepackage{multirow}
\usepackage{color,soul}
\usepackage{blkarray}
\usepackage{nicefrac}   
\usepackage[colorlinks,citecolor=blue,urlcolor=blue,filecolor=blue,backref=page]{hyperref}% compact symbols for 1/2, etc.
\usepackage{microtype}      % microtypography
\usepackage{lipsum}
\usepackage{graphicx}
\usepackage{anyfontsize}
\usepackage{afterpage}
\usepackage{adjustbox}
\usepackage{color, colortbl}
\usepackage{mathbbol}
\usepackage{mathrsfs}
\usepackage{hhline}
\usepackage{rotating}
\usepackage{comment}
\usepackage{tabularx}
\definecolor{Gray}{gray}{0.9}

\allowdisplaybreaks

\usepackage{arydshln}
\makeatletter
\renewcommand*\env@matrix[1][*\c@MaxMatrixCols c]{%
	\hskip -\arraycolsep
	\let\@ifnextchar\new@ifnextchar
	\array{#1}}
\makeatother

\RequirePackage{doi}

\renewcommand{\tilde}[1]{\widetilde{#1}}

\usepackage{bbm} % mathbbm
\newcommand{\bolds}[1]{\boldsymbol{#1}}

\newcommand{\calT}{{\cal T}}

\newcommand{\ba}{\bolds{a}}
\newcommand{\bA}{\bolds{A}}

\newcommand{\bB}{\bolds{B}}

\newcommand{\bC}{\bolds{C}}

\newcommand{\bD}{\bolds{D}}

\newcommand{\bg}{\bolds{g}}
\newcommand{\bG}{\bolds{G}}

\newcommand{\bI}{\bolds{I}}

\newcommand{\bP}{\bolds{P}}

\newcommand{\bs}{\bolds{s}}

\newcommand{\bV}{\bolds{V}}
\newcommand{\bw}{\bolds{w}}

\newcommand{\bx}{\bolds{x}}
\newcommand{\bX}{\bolds{X}}
\newcommand{\by}{\bolds{y}}
\newcommand{\bY}{\bolds{Y}}

\newcommand{\Trond}{\mathcal{T}}
\newcommand{\Ak}{\bA_k}
\newcommand{\Dk}{\bD_k}
\newcommand{\IdentityMat}{\bI}
\newcommand{\Norm}{\mathcal{N}}
\newcommand{\bzero}{\mathbf{0}}

\newcommand{\bbeta}{\bolds{\beta}}
\newcommand{\beps}{\bolds{\varepsilon}}
\newcommand{\btheta}{\bolds{\theta}}

\newcommand{\bgamma}{\bolds{\gamma}}

\newcommand{\bmu}{\bolds{\mu}}

\newcommand{\bLambda}{\bolds{\Lambda}}

\newcommand{\bOm}{\bolds{\Omega}}
\newcommand{\bpsi}{\bolds{\psi}}

\newcommand{\GP}{\mathcal{GP}}
\newcommand{\IG}{\mathcal{IG}}

\newcommand{\lrnd}{\left(}
\newcommand{\rrnd}{\right)}
\newcommand{\lsq}{\left[}
\newcommand{\rsq}{\right]}

\newcommand{\given}{\,|\,}

\newcommand{\tCollapsedurl}{\if0\blind
	{
		\url{\tCollapsedurl}
	} \fi
	\if1\blind
	{ 
		\url{github.com/}\texttt{[url redacted in blinded version]}
	} \fi}

\title{Bayesian Hierarchical Modeling and Analysis for Actigraph Data From Wearable Devices}

\author{
    Pierfrancesco Alaimo Di Loro\\
    \scriptsize{Dpt. GEPLI}\\
    \scriptsize{LUMSA}\\
    \scriptsize{\texttt{p.alaimodiloro@lumsa.it}}
\And
    Marco Mingione\\
    \scriptsize{Dpt. of Political Sciences}\\
    \scriptsize{Roma Tre University}\\
    \scriptsize{\texttt{marco.mingione@uniroma3.it}}
\And
    Jonah Lispsitt\\
    \scriptsize{Fielding School of Public Health}\\
    \scriptsize{University of California, Los Angeles}\\
    \scriptsize{\texttt{jonahlipsitt@gmail.com}}
\And
    Christina M. Batteate\\
    \scriptsize{Center of Occupational and Environmental Health}\\
    \scriptsize{University of California, Los Angeles}\\
    \scriptsize{\texttt{cbatteate@ucla.edu}}
\And
    Michael Jerrett \\
     \scriptsize{Fielding School of Public Health}\\
    \scriptsize{University of California, Los Angeles}\\
    \scriptsize{\texttt{mjerrett@ucla.edu}}
\And
    Sudipto Banerjee\\
     \scriptsize{Dpt. of Biostatistics}\\
    \scriptsize{University of California, Los Angeles}\\
    \scriptsize{\texttt{sudipto@ucla.edu}}
}

\begin{document}

\RestyleAlgo{boxruled}
	
	\def\spacingset#1{\renewcommand{\baselinestretch}%
		{#1}\small\normalsize} \spacingset{1}
		
\maketitle
\begin{abstract}
The majority of Americans fail to achieve recommended levels of physical activity, which leads to numerous preventable health problems such as diabetes, hypertension, and heart diseases. This has generated substantial interest in monitoring human activity to gear interventions toward environmental features that may relate to higher physical activity. Wearable devices, such as wrist-worn sensors that monitor gross motor activity (actigraph units) continuously record the activity levels of a subject, producing massive amounts of high-resolution measurements. Analyzing actigraph data needs to account for spatial and temporal information on trajectories or paths traversed by subjects wearing such devices. Inferential objectives include estimating a subject's physical activity levels along a given trajectory; identifying trajectories that are more likely to produce higher levels of physical activity for a given subject; and predicting expected levels of physical activity in any proposed new trajectory for a given set of health attributes. Here, we devise a Bayesian hierarchical modeling framework for spatial-temporal actigraphy data to deliver fully model-based inference on trajectories while accounting for subject-level health attributes and spatial-temporal dependencies. We undertake a comprehensive analysis of an original dataset from the Physical Activity through Sustainable Transport Approaches in Los Angeles (PASTA-LA) study to ascertain spatial zones and trajectories exhibiting significantly higher levels of physical activity while accounting for various sources of %\textcolor{purple}
{heterogeneity}. 
	\end{abstract}
	
	\noindent%
	{\it Keywords:} Bayesian hierarchical models, Directed acyclic graph, Gaussian processes, Physical activity, Sparsity, Spatial-temporal statistics
	%\vfill
	
	%\newpage
	%\spacingset{1.45}
	
\section{Introduction}\label{sec:intro} 
Promoting a healthy lifestyle continues to stoke substantial research activities in public health. The ``Physical Activity Guidelines for Americans'' (2nd edition) suggests that most individuals, depending on age and body composition, receive 150-300 minutes of moderate to vigorous physical activity (MVPA) weekly \citep{piercy2018physical}. In general, the scientific community agrees that regular physical activity (PA) can have immediate and long-term health benefits \citep{reiner2013long, bull2020world}. Despite these well-known benefits, most Americans fail to meet recommended requirements \citep{piercy2018physical}. Specifically, only 1 in 5 high-school adolescents and 1 in 4 adults meet recommended levels of physical activity. Given the well-established relationships between lack of PA and several obesity-related chronic conditions such as heart disease, type 2 diabetes, and cancer, as well as many physical and mental health benefits, an urgent need exists to improve monitoring of PA and to establish public health programs that promote more PA\footnote{More details at \url{https://www.cdc.gov/chronicdisease/resources/publications/factsheets/physical-activity.htm}}. 

Technologies for monitoring spatial energetics \citep{james2016, drewnowski2020} %(e.g., step rates, blood pressure, heartbeat, activity counts, etc.) 
and promoting physical activity continue to emerge. Among others, \emph{actigraphy} broadly refers to the monitoring of human rest and activity cycles using wearable devices. Actigraphy data are gathered directly from wearable sensors or indirectly through smart-phone mobile applications and record repeated measurements at a very high resolutions. In particular, accelerometers are motion sensors that measure acceleration along different axes and are able to collect large amounts of data \citep{plasqui2007physical, sikka2019analytics}. They are increasingly conspicuous because of their affordability, accuracy, and availability in smart-phones, smart-watches and other wearable devices. Many devices also include Global Positioning System (GPS) sensors that reference measurements with location tracking along trajectories, or paths, traversed by the subject. Collected data can be quickly downloaded and promptly analyzed to obtain insights into their pattern and structure. 

%In this paper, 
We pursue a comprehensive analysis of an original actigraphy data set from the Physical Activity through Sustainable Transport Approaches in Los Angeles (PASTA-LA) study. %Analyzing such data is sought for several reasons.
%\textcolor{purple}
{Actigraphy and GPS data analysis customarily involve \emph{idle} records that occur if a charged device does not detect acceleration over a specified time interval (e.g., 10 seconds). While idle records may correspond to periods of a subject's inactivity, they can also arise from other factors including technical malfunctions or the subject not wearing the device. On the other hand, eliminating idle records does not exclude all inactive periods because the accelerometer still records minor movements from where it is worn while a subject may be mostly inactive. Attempting to account for idle records as representative of inactive periods are likely to confound assessments of a subject's activity levels with current technological capabilities. Hence, we do not consider idle records and focus on the following specific data analytic aims:} (i) estimating a subject's physical activity levels along trajectories; (ii) identifying trajectories that are more likely to produce higher levels of physical activity for a given subject; and (iii) predicting expected levels of physical activity in any proposed new trajectory for a given set of health attributes. %(i) assessing health effects of different physical activity intensities \citep{pate2008evolving, smuck2014does, farooq2020longitudinal}; (ii) improving classification accuracy of activity intensity (e.g., sedentary, light, moderate, vigorous; \cite{degroote2020low, sagelv2020criterion}); and (iii) assessing the effectiveness of interventions and health promotion techniques \citep{troped2010built, dunton2014neighborhood, hartman2018patterns}.
Researchers 
%have cogently demonstrated the benefits of an active lifestyle over a sedentary one on physical and mental well-being and longevity \citep{lee1995exercise}.
%Therefore, researchers 
find actigraph tracking especially attractive as it allows for a better understanding of what behavioral and environmental factors influence population and individual health and, hence, aid in public health recommendation and policy. %Section~\ref{secData} provides details on the PASTA-LA data set.

Actigraphs generate data evolving over space and time, which suggests rich classes of space-time models for analysis \citep{gelfand2010handbook, cressie2015statistics}. In particular, actigraph analysis presents some notable challenges \citep{kestens2017}: the data sets are large, or even massive, as they are recorded at very high frequencies; they exhibit dependence along trajectories which should be accounted for both explanation and prediction \citep{ray2018physical, bai2018two}. We argue against a customary spatial-temporal process over $\mathbb{R}^2\times \mathbb{R}^{+}$ and propose disentangling spatial effects from temporal dependence along trajectories.    
The balance of the paper is organized as follows. Section~\ref{secData} introduces the PASTA-LA data-set with insights into accelerometry data. The model for the temporal correlation is introduced in Section~\ref{secMod}, while spatial effects are discussed in Section~\ref{sec:modSpatEff}. An extensive simulation study validating our model is proposed in Section~\ref{secExp}. Data analysis, model assessment and comparisons are presented in Section~\ref{sec:App}. Finally, we conclude with a discussion in Section~\ref{secDisc}.

\section{Data}\label{secData}
% \subsection{Data collection}\label{secDataCollection}
%\textcolor{purple}
{Our data set is compiled from the original \textbf{P}hysical \textbf{A}ctivity through \textbf{S}ustainable \textbf{T}ransport \textbf{A}pproaches in \textbf{L}os \textbf{A}ngeles (PASTA-LA) study conducted on a cohort of 460 individuals monitored between May 2017 and June 2018. Data were collected through different sources: online questionnaires, a smartphone app named \textit{MOVES}, a GPS device (GlobalSat DG-500), and a wearable actigraph unit (Actigraph GT3X+). Data collected through the \textit{MOVES} app, whose reliability must still be verified and discussed, are not considered in this paper.
While $460$ is the sample size of the complete study, the GPS and actigraph devices were deployed only on a nested sample of $134$ individuals due to cost considerations. We analyze data collected through these two devices that were supposed to be worn by the participants in the nested sample for two one-week periods (one in 2017 and one in 2018). Study protocols for safeguarding participant information received necessary institutional review board (IRB) approval. The data were stored on a secure computer and a redacted version was created for purposes of data sharing.} 
%While we do not pursue all of the aims of the PASTA-LA study,  we build and test the framework in Section \ref{secMod} for modeling high-frequency   actigraph data related to different individuals.

\subsection{Questionnaires}
% The online questionnaire was supposed to be repeated six times on the whole population, three times before the intervention (Bruin Bike Share Launch) and three times subsequently; this included two baseline surveys and four follow-up surveys. Each survey included a maximum of 184 variable responses per participant pertaining to the user's demographics and transportation habits. Not all participants completed all questionnaires, and the survey available on the largest part of participants is the \textit{First baseline questionnaire}, which contains mostly demographic information such as sex, age, BMI\footnote{binned according to the World Health organization (WHO) guidance (\url{https://apps.who.int/bmi/index.jsp?introPage=intro_3.html})}, ethnicity and other socioeconomic factors. Hence, we have considered only data resulting from this survey. 
The online questionnaires included two baseline and four follow-up surveys: one baseline and two follow-ups for each collection period of the actigraph and GPS data. Each survey consisted of responses pertaining to the participant's demographics and transportation habits. %Not all participants completed all questionnaires. Hence, we considered only the surveys available for all the participants. 
Here, we consider the \textit{first baseline questionnaire}, which is the only one available for all the participants in the nested sample. %\textcolor{purple}
{Personal information and other socioeconomic factors have been encoded as follows for subsequent analysis: 
\begin{itemize}
    \item \textbf{Sex}: Female or Male;
    \item \textbf{Ethnicity}: Asian, Black/African/Caribbean, Latin-American, White, or Other (mixed multiple ethnic groups or prefer not to answer);
    \item \textbf{Age (years) class}: $(0, 18]$, $(18, 25]$, $(25, 34]$, $(34, 45]$, $(45, 70]$;
    \item \textbf{BMI ($kg/m^2$) class}\footnote{according to standard guidelines of the Center of Disease Control and Prevention \url{https://www.cdc.gov/obesity/basics/adult-defining.html}}: underweight if $\mbox{BMI} \in (15, 18.5]$,  normal if $\mbox{BMI} \in (18.5, 24.5]$, overweight if $\mbox{BMI} \in (24.5, 30]$, and obese if $\mbox{BMI} >30$;
    \item \textbf{Yearly Income Level} (in thousands \$): $(0, 10]$, $(10, 25]$, $(25, 50]$, $(50, 75]$, $(75, 100]$, $(100, 150]$, $(150, +\infty]$, and Don't know/Prefer not to answer;
    \item \textbf{Educational attainment}: High-school diploma, College graduate, Associate degree, Graduate;
\end{itemize}
We filtered unreasonable values of the BMI, i.e. BMI$<10$, which was observed just for one individual, leaving 133 out of 134 individuals in the nested sample. A user ID was assigned to each survey response data and a redacted master key was generated using all ID types for joining with other study data.}

\subsection{Actigraph}\label{subsec:Actipreproc}
%The Actigraph units were were provided to a sample of $163$ individuals for two one-week periods%, before and after the Bruin Bike Share launch. 
%\textcolor{purple}
{The Actigraph unit is an accelerometer roughly the same size and weight of the average wrist-watch. It can be worn on the wrist, hip, and thigh and measure the directional acceleration at a specified time frequency (up to 100 Hz). The Actigraph GT3X+ model used for the PASTA-LA study can detect accelerations measured in gravitational units ($G$) with a sensitivity of $\pm 3$ milligravity ($mG$) in the three orthogonal planes (anteroposterior, mediolateral, and vertical).
Data are stored in an internal memory and can be downloaded to other hardware for analysis through a proprietary software. 
The participants were asked to start wearing the accelerometer on their dominant wrist as soon as it was handed to them, as the devices could have been properly calibrated at that time. The study protocol demanded that participants wear the Actigraph unit at all times other than during bathing and sleeping (awake time was assumed approximately from 7am to 11pm).  The sampling frequency has been set to 30 Hz and the \textit{idle sleep} mode has been activated in order to save battery and memory. With this mode on, the device would go idle every-time it records no acceleration ($<\pm 40mG$) for $10$ consecutive seconds.
The Actigraph GT3X+ %, differently from its predecessors, 
grants access to the a \texttt{.gt3x} file with the raw acceleration measurements. It can be loaded in \texttt{R} using the \texttt{read.gt3x} package and contains the raw accelerations at each timestamp.
Such accelerations comprise the basic ingredients to get a proxy for body movement from an accelerometer \citep{mathie2003detection, migueles2019comparability, bammann2021generation}. There are substantial investigations into its statistical relationships with PA measures, such as \textit{energy expenditure} measures (EE) \citep{crouter2006novel, freedson2012assessment, taraldsen2012physical} and the \textit{Metabolic Equivalent of Task} (MET)
\citep{lyden2014method, staudenmayer2015methods, migueles2017accelerometer, van2018wrist}.
Among various metrics, we take the instantaneous body vector Magnitude of Acceleration (MAG) as the primary endpoint of our analysis \citep{van2011estimation, white2016estimation, doherty2017large}. Further discussion about the conversion of MAG into energy expenditure measures is reported the Supplementary Material \citep{alaimo2023aoasSupplement}.}

%\textcolor{purple}
{We were able to retrieve the Actigraph raw data only on $K=97$ out of the $133$ original individuals.
Let $x$, $y$ and $z$ be the dynamic acceleration of the body of the $k$-th individual. The point-wise MAG is defined as:
\begin{equation}
\label{eq: MAG}
   \text{MAG}_{kt}=\sqrt{x_{kt}^2+y_{kt}^2+z_{kt}^2}, \qquad k=1,\dots,K.
\end{equation} 
%Participants reported forgetting to put it back on in the morning or not wanting to wear it in certain scenarios, also resulting in data loss. \cite{troiano2014evolution} showed that such a protocol naturally results in huge amounts of missing data, not random but biased toward an increased general level of physical activity (i.e. people who kept the accelerometer on during these times are likely to be the ones who would be performing physical activity). 
%When participants arrived at the research offices to drop off devices, some described issues of efficacy in the ability to keep the device on or charged. Indeed, while the actigraphs were supposed to hold a charge long enough to last the whole week, this was not always the case (possibly due to external conditions affecting the battery life or variations in manufacturing). This resulted in a large amount of missing data, which emphasize the need for an accurate and efficient strategy to interpolate the values over unobserved time windows. 
%During the download the data were aggregated in sample epochs spanning $10$ seconds. Measurements included the activity counts for the three axles and the step counts (obtained from the axles counts through a proprietary algorithm). 
%Time-stamps of the final measurements (hour, minute, and second) were referenced by the mid-point between the beginning and the end of the epoch.
However, the raw accelerations recorded by the accelerometer must be %properly cleaned and 
appropriately %
processed to glean body movement %from the gathered data 
\citep{doherty2017large}. Indeed, the raw acceleration recorded by each axle is the sum of both the \textit{static} and the \textit{dynamic} acceleration, but only the second is the effect of actual body movement.}
\begin{comment}
\vspace{0.5cm}
\textcolor{purple}{WE NEED TO DECIDE IF WE WANT TO MODEL ONLY THE ACTIVE TIME. THE TWO ALTERNATIVE SENTENCES MAY BE}

\textcolor{purple}{\begin{itemize}
    \item ALT 1: KEEP ALL. First of all, the \textit{idle mode} flag can be used to identify the non-wear time as long-lasting time of inactivity. Following other works on the topic, we set $15$ minutes of consecutive \textit{idle} records as a reasonable threshold to characterize non-wear time. Indeed, especially when accelerometers are worn on the wrist, it is particularly uncommon to not record any meaningful acceleration even when resting, sitting, or lying down.
    \item ALT 2: DROP IDLE. First of all, we drop all the \textit{idle} records, i.e. all the occasions in which the accelerometer recorded no acceleration for longer than $15$secs and the device went \textit{idle} (RESULTS NOW ON THE PAPER REFER TO THIS ALTERNATIVE).
\end{itemize}}
\end{comment}
%\textcolor{purple}
{First, we remove \textit{idle} records, i.e. all the occasions in which the accelerometer recorded zero acceleration for longer than $15$ seconds and the device went \textit{idle}. %This is consistent with our analytic goals of estimating MAG levels when the subject is active. 
It is very unlikely that these idle records with zero acceleration coincide with a subject's inactive periods because the accelerometer still records positive, albeit small, magnitudes of accelerations over inactive periods due to movements in the wrists, hips and thighs. Idle records, on the other hand, are likely to arise from technical malfunctions or from a subject violating protocol and not wearing the device in the experimental time window.}

%\textcolor{purple}
{Second, the raw accelerations recorded by single axles must be disentangled from unwanted static or non-static components: the effect of the earth's gravitational force and other external accelerations (e.g car, bus, elevators) at low frequencies, machine noise and vibrations at high frequencies. To address this issue, we adopted a Band-Pass Butterworth digital filter of order $4$ with frequency window $(0.25, 10)$ to clean the signals from these long and short waves \citep{mathie2003detection}. Indeed, most human activities result in signals with a frequency between $0.25$Hz and $10$Hz \citep{khusainov2013real}. An example of how the raw signal is modified through this process is provided in the Supplementary Material \citep{alaimo2023aoasSupplement}. We subsequently evaluate the point-wise $\text{MAG}_{kt}$ using the filtered accelerations $\lrnd\tilde{x}_{kt}, \tilde{y}_{kt}, \tilde{z}_{kt}\rrnd$. However, the instantaneous MAG evaluated at the original $30$Hz frequency is extremely erratic and the single value may not represent well the PA intensity of the participant at that time. For this reason, it is usually averaged over $5$ to $10$ second epochs to acquire a more suitable measure of PA intensity \citep{migueles2017accelerometer, doherty2017large}. Here, we perform a kernel smoothing of the $30$Hz measurements in order to be representative of the single time-point, and get a 1 second time resolution. The resulting vector magnitude is
\begin{equation}
\label{eq:magsmooth}
    \tilde{\text{MAG}}_{kt} = \sum_{j}k_b(t-t_j)\cdot \text{MAG}_{kt_j},
\end{equation}
where $k_b(\cdot)$ is a Gaussian kernel with bandwidth $b=5$ seconds. This ensures that the impact of the neighboring points becomes negligible for $|t-t_j|>10$ seconds.
Finally, we removed all the observations recorded outside of the pre-specified daily time-window, i.e. from 7am to 11pm.}

\subsection{GPS}
%The \textit{Global Positioning System} (GPS) is a satellite-based radio-navigation system that does not require the user to transmit data and operates independently of any telephonic or internet reception. Any GPS unit can be set to record and store the spatial location at a pre-specified time frequency so that they could be downloaded and subsequently analyzed in a second moment. Obstacles, such as mountains and buildings, can block the relatively weak GPS signals and prevent the device from functioning accurately. 

%\textcolor{purple}
{The GPS device \textit{GlobalSat DG-500} recorded the subject's location (latitude and longitude) roughly every $5$ seconds, together with date, time, and speed ($km/h$, measured as distance over time through linear interpolation). 
This work restricts the attention to 93 out of 97 subjects living and working in the Westwood neighborhood of Los Angeles in order to avoid a geographical imbalance that could bias and invalidate the model estimates. This area hosts the university campus of UCLA and it includes the largest part of all the available observations. Westwood is a walk-friendly neighborhood with a lot of green areas, parks, and major roads with shops and amenities. People were free to move inside and between buildings (e.g. people at the gym, office, etc) and we are interested in quantifying their movement in all these settings.}

%\textcolor{purple}
{However, GPS measurements can be affected by possible inaccuracies, especially around buildings, that may cause unreasonable jumps in a very small time-span. We note that most of these issues are already mitigated by an automatic filtering process of the GPS device, that would drop records for which the signal is not strong enough. Nevertheless, to further enhance the cleaning process of GPS measurements, we removed all data points for which the computed average speed between two subsequent points was larger than $100km/h$. We picked such a high threshold as we do not want to drop observations related to individuals standing or sitting in a bus or car.}

\subsection{External covariates}\label{subsec: covariates}
\begin{figure}[H]
    \centering
    \begin{subfigure}[b]{0.32\textwidth}
    \includegraphics[width=.95\textwidth]{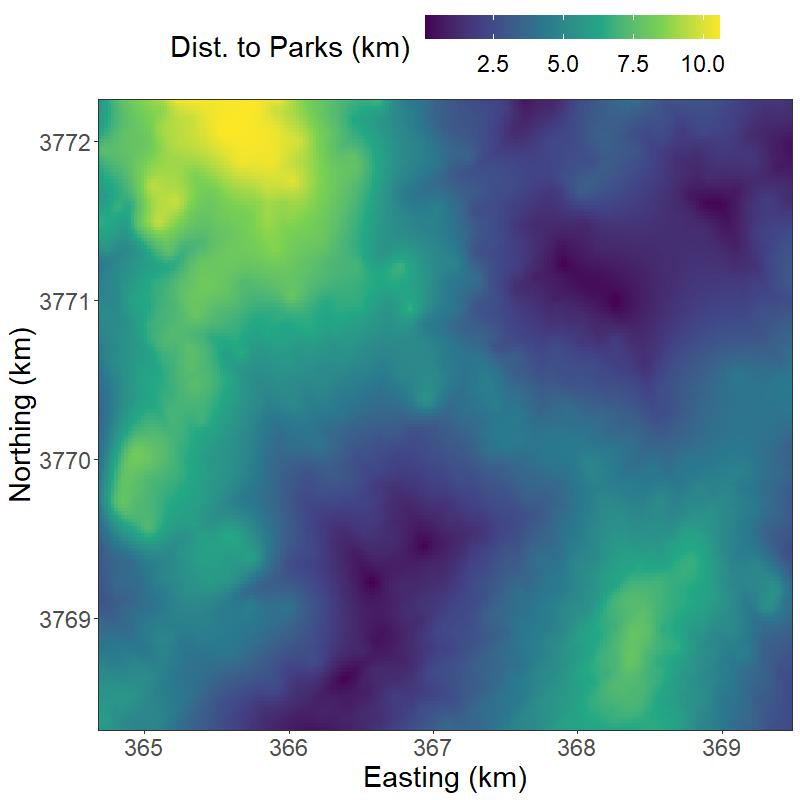}
    \caption{}
    \label{dtp}
    \end{subfigure}
    \begin{subfigure}[b]{0.32\textwidth}
    \includegraphics[width=.95\textwidth]{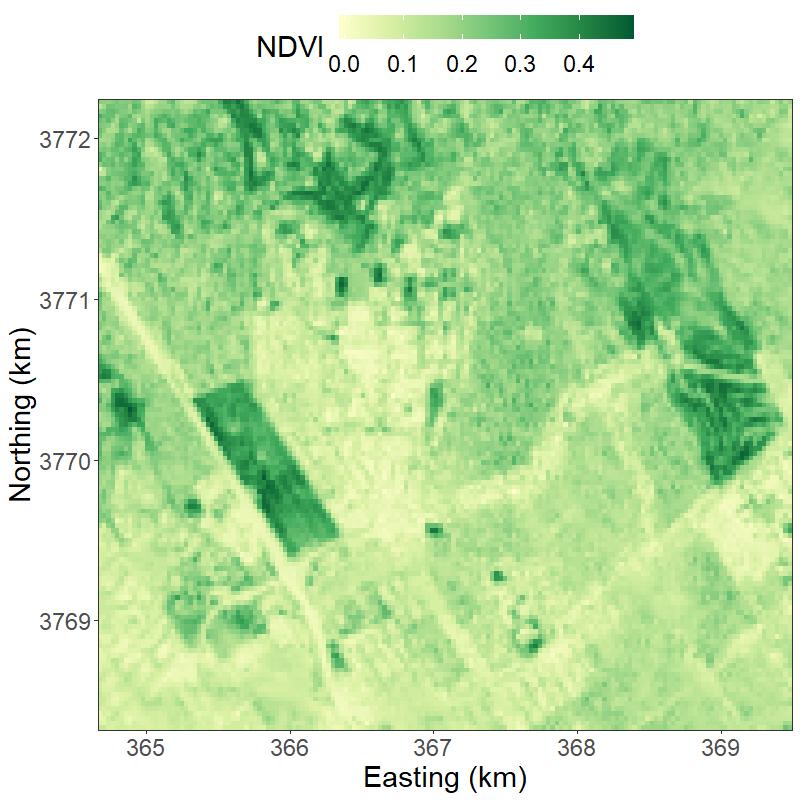}
    \caption{}
    \label{ndvi}
    \end{subfigure}
    \begin{subfigure}[b]{0.32\textwidth}
    \includegraphics[width=.95\textwidth]{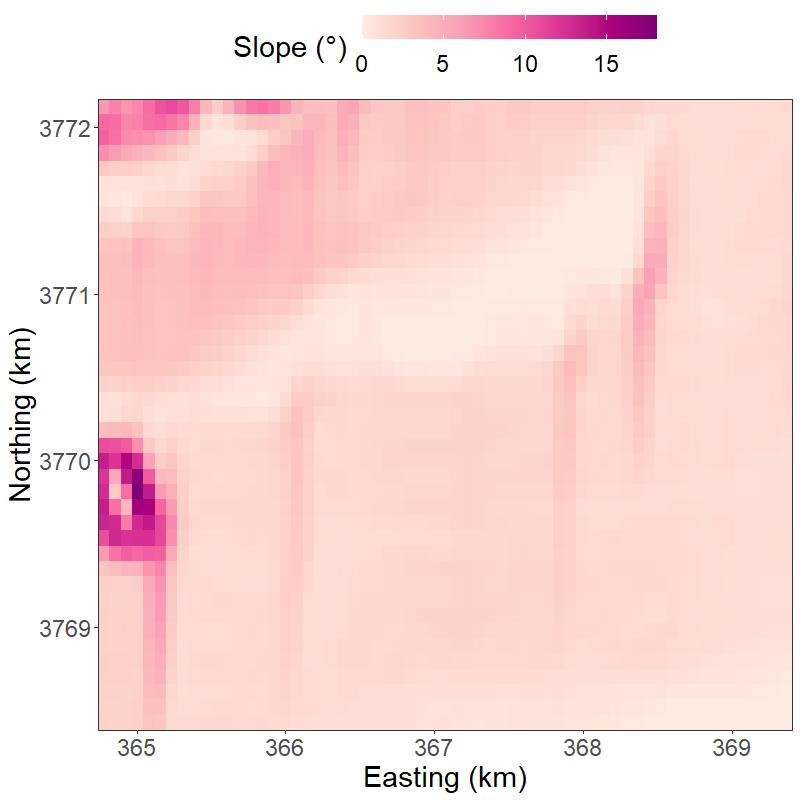}
    \caption{}
    \label{slope}
    \end{subfigure}
    \caption{Spatial-varying covariates: (a) distance to parks, (b) NDVI and (c) Slope.}
    \label{fig:extcov}
\end{figure}

%\textcolor{purple}
{PA levels are not only affected by individual characteristics, but can be fostered by specific features of the surrounding area they are navigating. Therefore, we included three external covariates to account for some of the built-in environment features of the Westwood area. In particular, we used the following.
\begin{itemize}
    \item[(i)] The weighted overlay distance to parks (in $km$) with a spatial resolution of $23\times 23$ which can be downloaded from \url{https://egis2.lacounty.gov/arcgis}. It represents a weighted distance of each point from officially recognized parks and it can be seen as a proxy of the \textit{green area density} (see Figure \ref{dtp}; darker shades indicate proximity to parks). %(ASK JONAH FOR MORE DETAILS)
    \item[(ii)] The Normalized Difference Vegetation Index (NDVI), which is available with a spatial resolution of $30\times 30$, can be downloaded from \url{https://earthexplorer.usgs.gov/} and provides a measure of the \textit{greeness} of the patch itself (see Figure \ref{ndvi}; darker shades depict more greenness).
    \item[(iii)] The slope (azimuth), with a spatial resolution of $23\times 23$, is computed from the digital elevation model (DEM) downloaded from \url{http://www.webgis.com/terr_pages/CA/dem1/losangeles.html}. It represents the average angular inclination of the ground patch with respect to the horizon line (see Figure \ref{slope}; darker shades depict higher slopes).
\end{itemize}
While previous studies \citep[e.g.,][]{maddison2009} 
have reported on these variables affecting PA levels%. However
, they usually consider the environmental impact on the average PA level through a buffer around the home location of the participant, and not %as having a direct effect 
on its instantaneous PA level. With our current work, we want to discover and establish direct associations between PA levels of a subject and these spatially-indexed covariates along trajectories.}

\subsection{Joining}
\begin{figure}[ht]
    \centering
    \includegraphics[width=.7\textwidth]{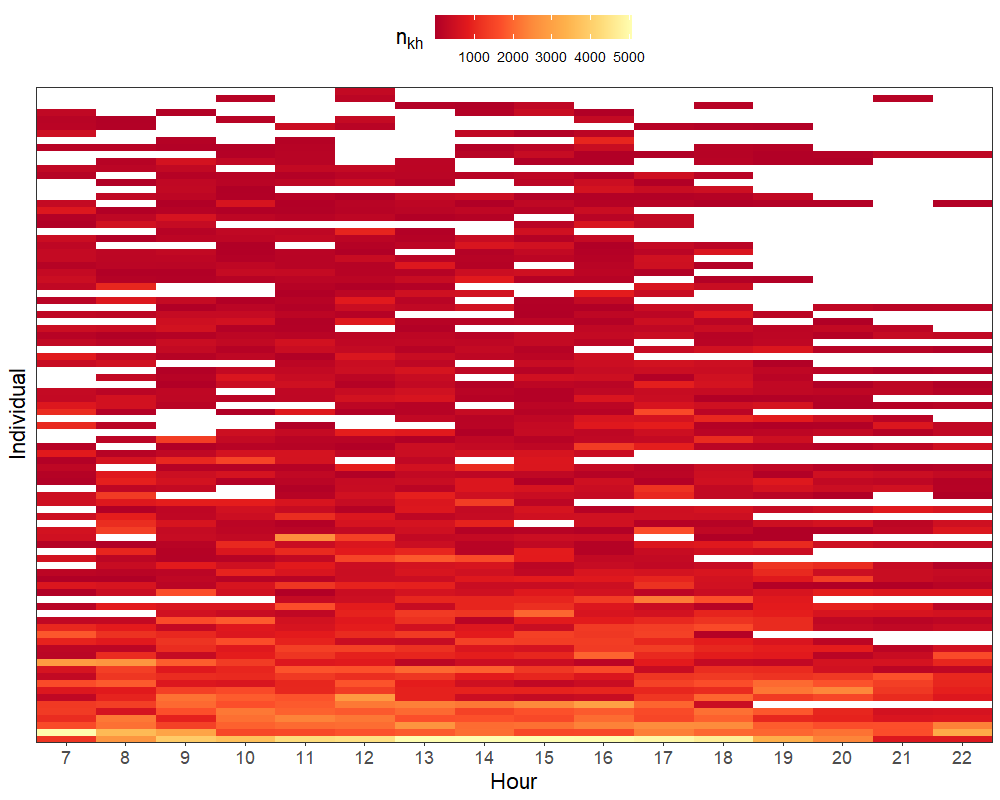}
    \caption{Missing data pattern by hour of the day and individual.}
    \label{fig:misspattern}
\end{figure}

\begin{figure}
\centering
%\begin{subfigure}[b]{.5\linewidth}
  \centering
  \includegraphics[width=.5\textwidth]{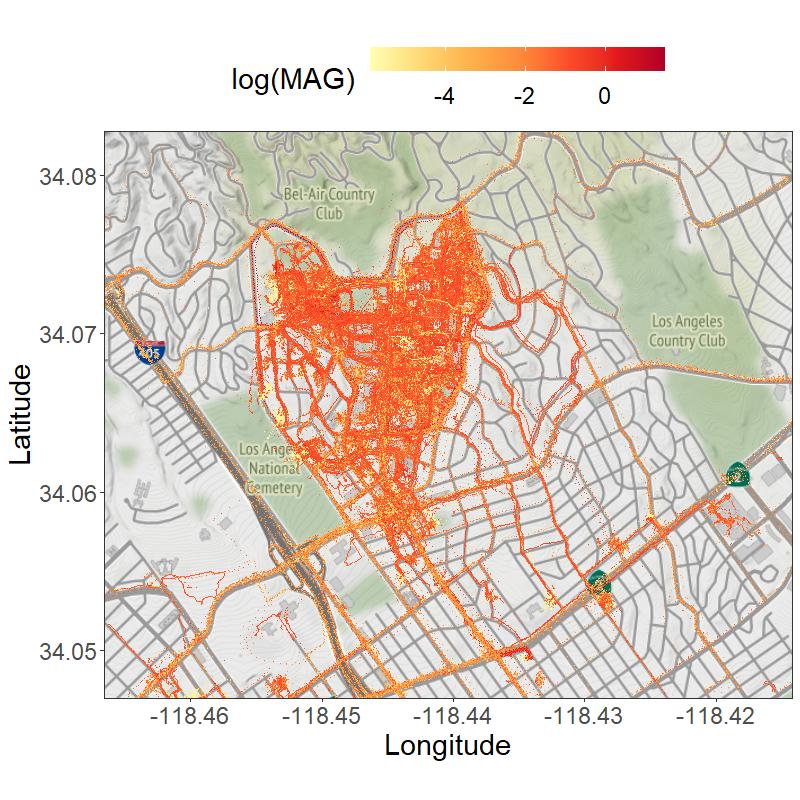}
  %\caption{}
     %\label{fig:obsknotsonWW}
%\end{subfigure}%
% \begin{subfigure}[b]{.5\linewidth}
%      \centering
%     \includegraphics[width=\textwidth]{images/MAGWestwoodMapFirst10PIDs.pdf}
%     \caption{}
%     \label{fig:MAGonWW}
% \end{subfigure}
\caption{Observed locations over the Westwood area. %(b): Observed physical activity level over the Westwood area on a subset of $10$ individuals
}
\label{ObsWW}
\end{figure}

%RStudio and the \texttt{tidyverse} package were used to convert the resulting data for usage in R.
GPS and accelerometer data were all assigned a participant ID aligned with the questionnaires' master-key to facilitate joining across all ID types (including email) while redacting and encrypting user data. The first baseline questionnaire, Actigraph, and GPS were available for the aforementioned group of $K=93$ individuals. Henceforth, we refer to this specific group of units.
%\textcolor{purple}
{The joining of different data sources follows these steps.
\begin{enumerate}
    \item Actigraph data are joined to the first baseline questionnaire using the individual master-key. The resulting data set includes the physical activity endpoint recorded by the Actigraph at the different timestamps and all the available individual information, but no spatial information.
    \item GPS data are joined to the dataset obtained in 1) using the individual master-key and the timestamp. Following the processing in Section~\ref{subsec:Actipreproc}, the Actigraph data are available at the same time resolution of GPS data as long as the subject wore them simultaneously% (but not vice versa)
    . Therefore, we decided to use the GPS as the leading table in the joining process. This avoids %filling our data set with
    use of %
    artificial data (e.g. interpolating GPS locations)% that could be far from reality
    .
    \item Spatial covariates are joined to the data set obtained in 2) through the \textit{minimum distance} criteria, i.e. each location is assigned the value of the closest point on the grid for each spatial-varying covariate.
\end{enumerate}}
%\textcolor{purple}
{The temporal coverage is not balanced across individuals because (i) subjects move around Westwood in different segments of the overall time window; and (ii) some participants violate the study protocol. Indeed, not all the participants were available for both of the one-week surveillance periods in $2017$ and $2018$. In fact, only $58$ out of the $93$ participants have data for the first week only and missed the follow-up survey. In the end, we go from the least represented individuals having $\approx 5\times 10^2$ observations ($\approx 2$ hours of data) to the most represented ones with more than $\approx 5 \times 10^5$ observations ($\approx 14$ days of data). Considering all the $5$ second time segments between the first and last observed point of each individual in each day as the potential observation window, the proportion of missing measurements ranges between $\approx 31\%$ and $\approx 97\%$. The overall proportion of missing measurements in the entire database (based upon aggregated $5$-second time segments for all individuals) is $\approx 83\%$. Figure~\ref{fig:misspattern} shows the number of observations for each individual $k = 1, \dots, K$ in each hour $h = 7, \dots, 22$, where the y-axis has been ordered in ascending order according to proportion of missing measurements for each subject. Overall, we can state that all hours are well-represented, but only few individuals have data for the whole daily time window. Summing up, all subsequent analysis will refer to the final data set consisting of $N \simeq 7 \times 10^5$ measurements across $K=93$ individuals, scattered over Westwood (see Figure~\ref{ObsWW}).}

\section{The model}\label{secMod}
%We devise a class of models to analyze high-frequency temporal observations belonging to different individuals. The proposed model takes into account the dependence structure among different realizations belonging to the same individual. Computations emanating from the massive dataset are addressed using efficient sparse matrix multiplication and inversion.

The outcomes corresponding to the $K = 93$ subjects are referenced with respect to the time at which they are recorded and the position in the trajectory. While it is tempting to work with a spatiotemporal process, dependence introduced by such processes may not be appropriate. An individual can visit the same location numerous times in his/her trajectory. These revisits need not occur at regular intervals and can be at distant time points. This suggests that proximity of two spatial locations in a trajectory need not result in strongly dependent MAGs recorded there. It appears more reasonable to model dependence among MAG measurements through a temporal process. In fact, such temporal processes can be motivated by the position vectors defining the trajectories as we describe below.   

\begin{figure}[t]
    \centering
    \begin{subfigure}[b]{0.45\linewidth}
    \centering
    \includegraphics[width=0.8\linewidth]{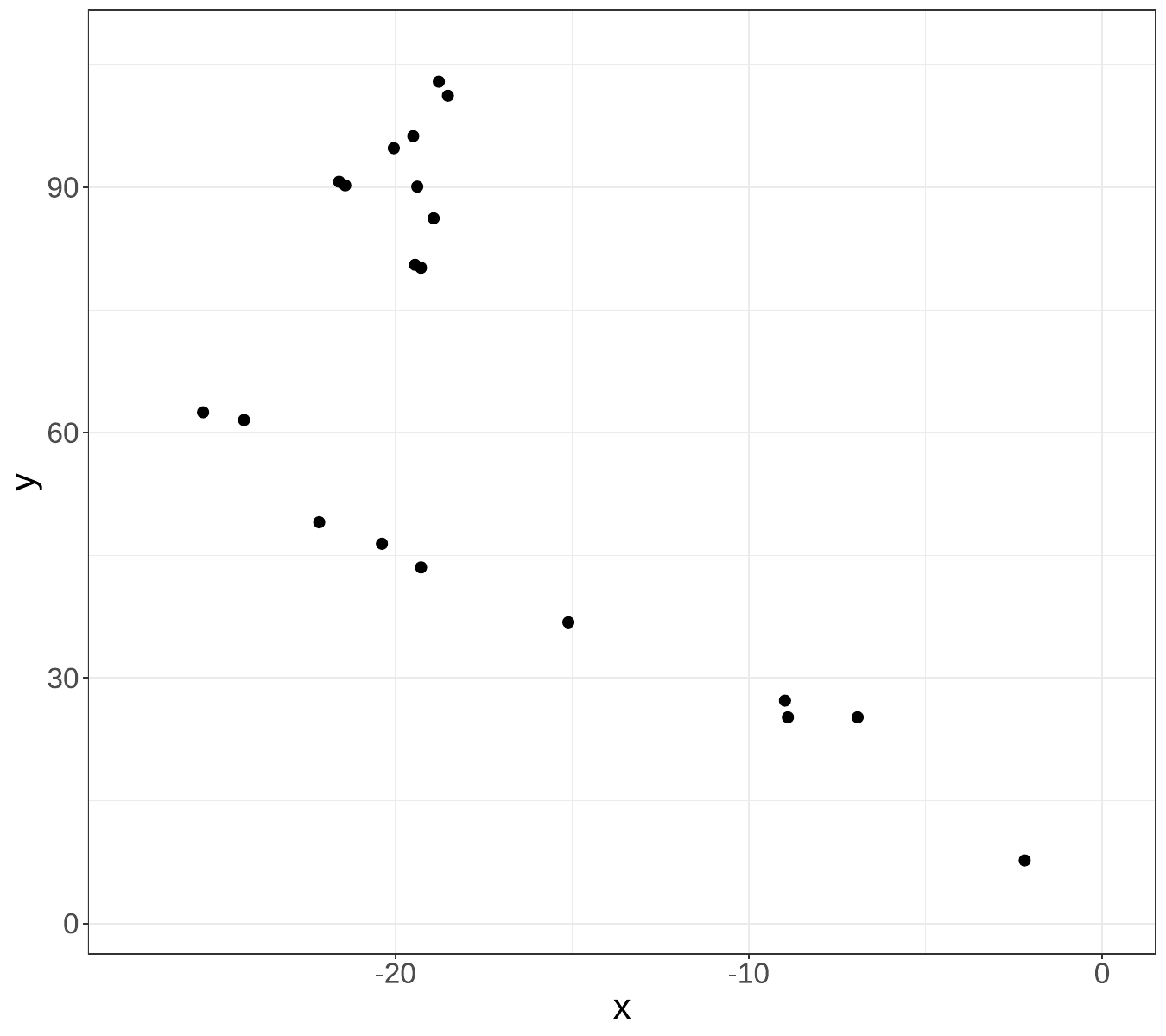}
    \caption{}
    \label{pointsjust}
    \end{subfigure}
    \begin{subfigure}[b]{0.45\linewidth}
    \centering
    \includegraphics[width=0.8\linewidth]{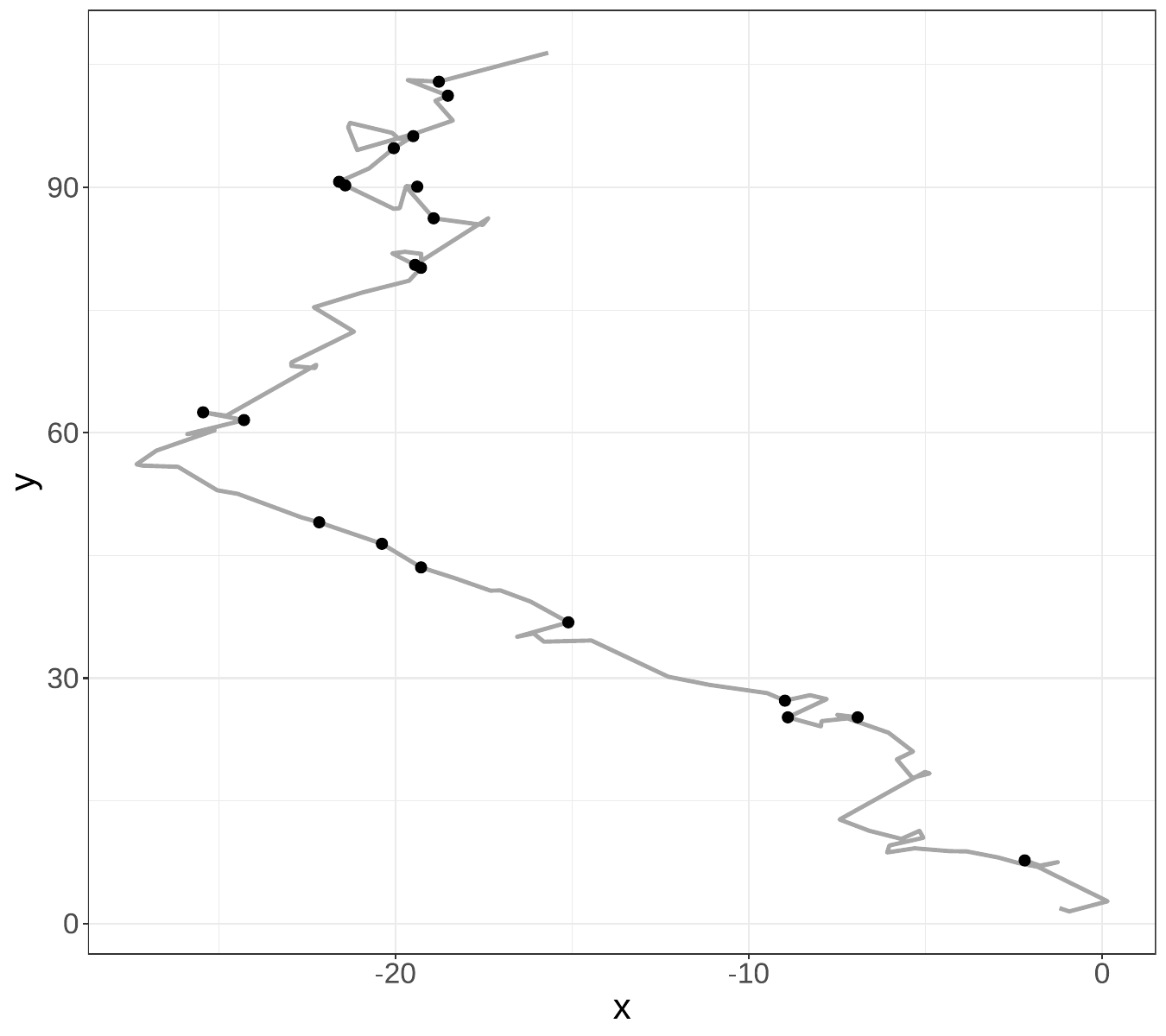}
    \caption{}
    \label{pathjust}
    \end{subfigure}
    \caption{Example of observed points (a) and trajectory (b): black dots are realizations, grey line is domain of the process}
    \label{fig:trajEx}
\end{figure}

Let $Z_k(\cdot):\mathbb{R}^2\rightarrow \mathbb{R}$ be a spatial process corresponding to individual $k$. The domain of $Z_k(\cdot)$ is restricted to the trajectories $\gamma_k(t)=\left(\gamma^x_k(t),\gamma^y_k(t)\right)$, where $k=1,\dots,K$ and $t\in\mathbb{R}^+$, which defines the movements of the $k$-th individual along time. As shown in Figure \ref{fig:trajEx}, the process actually belongs to a one-dimensional space, for which we define a proper distance measure $d(t_{ki}, t_{kj})=\|\gamma_k(t_{kj})-\gamma_k(t_{ki})\|$, where $t_{ki}$ is the $i$-th recorded time point from individual $k$. We approximate such distances as the elapsed time between the two points $d(t_{ki}, t_{kj}) = |t_{kj}-t_{ki}|$, which would result in a good approximation of the spatial distance (especially if the subject is moving at constant speed). More generally, the elapsed separation across time will reflect dependence better than the spatial distance. The faster an individual is moving from one point to the other, the shorter the time elapsed, and higher the correlation between the two measurements. Hence, we model our measurements as $Y_k(\cdot) \equiv Z_k\circ \gamma_k (\cdot) : \mathbb{R}^{+}\rightarrow\mathbb{R}$%Hence, we decide to parametrize the process just in terms of the time $t$
, 
%\begin{equation*}
%Z_k\circ \gamma_k (\cdot)\equiv Y_k(\cdot):\mathbb{R}^+\rightarrow\mathbb{R},
%\end{equation*}
which, by construction, is %proved to be 
a valid stochastic process% in \cite{abdalla2018coastline}
.

This will form the edifice of the model in Section~\ref{sec:tempMod}, where we are modeling the dependence by solely considering stochastic evolution through time. How should spatial information be introduced in the model? 
%does not mean that we must neglect spatial information at all. Indeed, we may still assume that the location $s_{ki}=\gamma_k(t_{ki})$ where the $i-th$ observation of the $k$-th individual is recorded has an effect on the outcome $Y_k(t_{ki})$. One may argue in favor of considering a spatio-temporal process, but this solution does not fit the nature of our data. 
%While individuals share common underlying spatial factors, 
Two individuals at the same spatial coordinate experience the same spatial effect but different temporal effects because their physical activities are a function of their trajectory's temporal evolution. An added complication is that trajectories intersect and overlap and, in practice, can have multiple observations at the same location. Even more flexible spatiotemporal covariance kernels (e.g., nonseparable or nonstationarity) will struggle to recognize the above features. Hence,
%Therefore, the spatial effect must act independently from the intrinsic temporal individual process. However, we cannot naively consider an additional spatial process, since  Therefore, in order to not over-complicate the dependence structure of the observed process, 
we introduce the spatial effect in the mean using \textit{spline regression} (see Section \ref{sec:modSpatEff}). 

\subsection{Temporal model}
\label{sec:tempMod}
Let $\Trond=\cup_{k=1}^K\Trond_k$ where $\Trond_k=\left\lbrace t_{ki}\right\rbrace_{i=1}^{T_k}$ and $t_{ki}\in\mathbb{R}^+$ be the set of the $n=\sum_{k=1}^K T_K$ observed time points. We model $\bY(\Trond)$ as the finite realization of a $K$-variate process $\bY(\cdot)$ over $\mathbb{R}^+$:
\begin{equation}
\label{eq:tempMod1}
    \bY(t) = \bX(t, \bgamma(t))^{\top}\bbeta + \bw(t) + \beps(t), \quad t\in\mathbb{R}^+,
\end{equation}
where $\bY(t) = (Y_1(t), Y_2(t),\ldots, Y_K(t))^{\top}$ is a $K\times 1$ vector of measurements at time $t$ on the $K$ individuals, $\bX(t,\bgamma(t))$ is a $p\times K$ matrix, each row being the values of a covariate for the $K$ individuals, $\bw(t)=(w_1(t), w_2(t),\ldots,w_K(t))^{\top}$ is a $K\times 1$ vector comprising a temporal process for each individual, and $\beps(t)\sim \Norm_K(0, \tau^2\IdentityMat_K),\; \tau^2\in\mathbb{R}^+,$ is a white noise process for measurement error. Each element of $\bw(t)$ is specified as $w_k(t)\stackrel{ind}{\sim} \GP\left(0, c_{\btheta}(\cdot,\cdot)\right)$, where $c_{\btheta}(\cdot,\cdot)$ is a  covariance function with parameters $\btheta\in\Theta$.

Let $y_{ki}$ and $\bx_{ki}$ be the outcome and covariates for individual $k$ at time point $t_{ki}$, respectively, so $\left\lbrace\left(y_{ki}, \bx_{ki}\right): \; k=1,\dots,K, \; i=1,\dots T_k\right\rbrace$ is the observed data. Let $\by_k$ and $\bw_k$ be $T_k\times 1$ vectors comprising all measurements and random effects on patient $k$, respectively. Forming the $n\times 1$ vectors
$\by=\begin{bmatrix}
\by_{1\cdot}^\top &
\by_{2\cdot}^\top &
\cdots &
\by_{K\cdot}^\top
\end{bmatrix}^\top$ and $\bw=\begin{bmatrix}
\bw_{1\cdot}^\top &
\bw_{2\cdot}^\top &
\cdots &
\bw_{K\cdot}^\top
\end{bmatrix}^\top$, and the $n\times p$ matrix $\bX=\begin{bmatrix}
\bX_{1\cdot}^\top&
\bX_{2\cdot}^\top&
\ldots &
\bX_{K\cdot}^\top
\end{bmatrix}^\top$, where $\bX_k$ is the $T_k\times p$ matrix of predictors corresponding to $\by_k$, we extend (\ref{eq:tempMod1}) to a hierarchical model with posterior distribution
\begin{equation}\label{eq:hierModel}
%\by = \bX\bbeta + \bw + \beps,\; \bw\sim \Norm_n\left(0, \bC_\theta\right),\; \beps\sim \Norm_n\left(0, \tau^2\cdot\IdentityMat_n\right)\;,
p(\bbeta, \bw, \btheta, \tau^2\given \by) \propto p(\btheta,\tau^2)\times N(\bbeta\given \bmu_{\beta}, \bV_{\beta})\times N(\bw\given \bzero, \bC_{\btheta}) \times N(\by\given \bX\bbeta + \bw, \tau^2\bI_n)\;.
\end{equation}
The covariance matrix $\bC_{\btheta} = \mbox{diag}\left(\bC_{\btheta_1,1},\bC_{\btheta_2,2},\ldots,\bC_{\btheta_K,K}\right)$ is $n\times n$ block-diagonal with $\bC_{\btheta_k,k}=[c_{\btheta}(t_{ki},t_{kj})]$ as the $T_k\times T_k$ temporal covariance matrix corresponding to individual $k$. Each individual is allowed its own covariance parameters, $\btheta_k$, and $\btheta = \{\btheta_1,\btheta_2,\ldots,\btheta_K\}$ in (\ref{eq:hierModel}) is the collection of all the covariance kernel parameters. 
\begin{comment}
\begin{equation*}
    \left[\bC_\btheta\right]_{ij} =c^*_\btheta(t_{kp}, t_{lq})=\begin{cases}
        c_\btheta(t_{kp}, t_{lq})  & k=l\\
        0 & k\neq l
    \end{cases}.
\end{equation*}
%In what follows, we will refer to $\bC_{\btheta}$ as $\bC$.

Full inference from (\ref{eq:hierModel})  straightforwardly set up by ascribing to the set of parameters $(\bbeta, \btheta)$ a suitable prior $\pi(\bbeta,\btheta)$ and deriving the posterior distribution as:
\begin{equation*}
    \pi(\bbeta,\btheta|\by)\propto \Norm_n\left(\by|\bX\bbeta, \tau^2\cdot\IdentityMat_n\right)\times\Norm_n\left(\bw|0, \bC\right)\times\pi(\bbeta,\btheta).
\end{equation*}
While this solution looks complete and elegant, its application
\end{comment}
Applying (\ref{eq:hierModel}) involves the determinant and inverse of $\bC_{\btheta}$, which require $\mathcal{O}(n^2)$ storage space and $\mathcal{O}(n^3)$ floating point operations (flops). The block-diagonal structure of $\bC_{\btheta}$ considerably alleviates this burden since $\mbox{det}(\bC) = \prod_{k=1}^K\mbox{det}(\bC_{\btheta_k,k})$ and $\bC^{-1} = \mbox{diag}\left(\bC_{\btheta_1,1}^{-1},\bC_{\btheta_2,2}^{-1},\ldots,\bC_{\btheta_K,K}^{-1}\right)$.
%This makes the estimation of such a model unfeasible for the majority of contemporary applications, and in particular for the data in consideration that include more than $10^6$ observations. However, our set of data has characteristics that may ease this troublesome computational burden.Indeed, the assumption of independence across individuals set to $0$ the covariance between observations belonging to different units, notwithstanding their distance along time.Therefore, the joint distribution of vector $\bw$, built by stacking the $\bw_k$ all together, has a covariance matrix with a block-diagonal structure $\bC=\oplus_{k=1}^K\bC_k$. Recalling that $T_k$ is the number of time-points for each individual, each block will be of size $T_k\times T_k$. In case of block-diagonal matrices, the inverse can be easily computed by inverting independently each sub-matrix on the diagonal and the determinant can be obtained by multiplying the determinants of the all the sub-matrices:
This reduces the flop count from $\mathcal{O}(n^3)=\mathcal{O}((\sum_{k=1}^KT_k)^3)$ to $\mathcal{O}(K\sum_{k=1}^K(T_k)^3)$, with a significant saving of calculations especially when the $T_k$'s are reasonably small ($<10^4$). Furthermore, each $\bC_{\btheta_k,k}$ can be computed in parallel rendering further further scalability to the algorithm.

%\footnote{Notice that, even if we are assuming the individual latent effects $w_k(\cdot)$ to evolve independently over time, we are still pooling the information coming from all the individuals for the estimation of the common vector of coefficients $\bbeta$ and covariance parameters $\btheta$.}

However, analyzing the Actigraph data in Section~\ref{secData} will involve $T_k > 10^5$ measurements from some individuals. Full inference will be impractical without any exploitable structure for each $\bC_{\btheta_k,k}$. Analyzing massive spatiotemporal data has witnessed burgeoning interest and a comprehensive review is beyond the scope of this work \citep[see, e.g.,][and references therein]{banerjee2017high, heaton2019case}. We will pursue an approximation due to Vecchia \citep{vecchia1988estimation} that has generated substantial recent interest \citep{datta2016hierarchical, datta16b, katzfuss2020, katzfuss2021, peruzzi2022highly} in scalable Bayesian modeling.

\subsection{Independent DAG models over individuals}\label{secindNNGP}
We adapt \textit{Vecchia's} likelihood approximation \citep{vecchia1988estimation} to the random effects $\bw_k$ for each $k=1,2,\ldots,K$. Beginning with the observed time points $\{t_{k1}< t_{k2} < \cdots < t_{kT_k}\}$ for individual $k$ and the directed acyclic graphical (DAG) representation $p(\bw_k)= p(w_{k1})\prod_{i=2}^{T_k} p(w_{ki}\given w_{k1},\ldots,w_{k(i-1)})$, we define  
\begin{equation}\label{eq:vecchiaApp}
p(\bw_k)\approx \tilde{p}(\bw_k)= p(w_{k1})\prod_{i=2}^{T_k} p(w_{ki}|\bw_{k,N(i)})\;,
\end{equation}
where $\tilde{p}(\cdot)$ is the joint density derived from $p(\bw_k)$ by restricting the parents (conditional sets) of each $w_{ki}$ in the DAG to a set $w_{kN(i)} = \{w_{kj} : j\in N(i)\}$, where $N(i)$ is a set of prefixed size $m$ comprising  the $m$ nearest neighbors of $t_{ki}$ from the past. Thus, $N(i) = \{t_{k(i-m)}, \ldots, < t_{k(i-1)}\}$ for $i>m$ and $N(i)=\{t_{k1},\ldots,t_{k(i-1)}\}$ for $i \leq m$. Such approximations yield valid probability likelihoods \citep{lauritzen1996graphical, stein2004approximating, murphy2012machine} and can be extended to stochastic processes \citep{datta2016hierarchical} for inference on arbitrary time points.

The connection between sparsity and conditional independence follows by writing (\ref{eq:vecchiaApp}) as a linear model $\bw_k=\Ak\bw_k+\boldsymbol{\eta}_k$,
\begin{comment}
\begin{equation}
\begin{aligned}
    &\bw_k=\Ak\bw_k+\boldsymbol{\eta}_k,\\
    &\boldsymbol{\eta}_k\sim\Norm_{T_k}(0,\Dk)
    \end{aligned}
    \label{adEquation}
\end{equation}
\end{comment}
where $\Ak$ is a $T_k\times T_k$ strictly lower triangular matrix, $\boldsymbol{\eta}_k\sim\Norm_{T_k}(\bzero,\Dk)$ and $\Dk$ is the $T_k\times T_k$ diagonal matrix such that $\left[\Dk\right]_{ii}=d_{ii}=\text{Var}\left(w_{ki}|\lbrace w_{kj}, j<i\rbrace\right)$ for $i=1,\dots,T_k$. The DAG imposes the lower-triangular structure on $\Ak$ and its $(i,j)$-th entry is allowed to be nonzero only for $j\in N(i)$.   
Therefore, each row of $\Ak$ has at most $m$ nonzero entries so that $\tilde{\bC}_k^{-1}=(\IdentityMat_{T_k}-\Ak)^\top\Dk^{-1}(\IdentityMat_{T_k}-\Ak)$ %identifies with the Cholesky decomposition of $\bC_k^{-1}$.
is sparse, where $\tilde{\bC}_k^{-1}$ is the precision matrix corresponding to $\tilde{p}(\bw_k)$. Replacing $\bC$ with $\tilde{\bC}$ in (\ref{eq:hierModel}) yields a computationally efficient hierarchical model with $N(\prod_{k=1}^K N(\bw_k\given \bzero, \tilde{\bC}_k)$ as the prior on $\bw$. 

The key observation is that the nonzero elements of the $i$-th row of $\Ak$ is the solution $\ba_k$ of the $m\times m$ linear system $\bC_{\btheta,k}[N(i), N(i)]\ba_k = \bC_{\btheta,k}[N(i),i]$, where $[\cdot,\cdot]$ indicates submatrices defined by the given row and column index sets. Obtaining the nonzero elements of $\Ak$ and $\Dk$ costs $\mathcal{O}(T_km^3)$ (scales linearly with $T_k$) instead of $\mathcal{O}(T_k^3)$ as would have been without sparsity. This cheaply delivers the quadratic form $\bw_k^\top\tilde{\bC}_k^{-1}\bw_k$ in terms of $\Ak$ and $\Dk$ and the determinant $\text{det}(\tilde{\bC}_k)=\prod_{i=1}^{T_k}d_{ii}$ at almost no additional cost. The lower triangular matrix $\Ak$ is not just sparse but also banded, with a lower bandwidth equal to $m$. Consequently, $\tilde{\bC}_k^{-1}$ is also banded with lower and upper bandwidth equal to $m$. This leads to further accrual of computational benefits. The overall cost is $\mathcal{O}(\sum_{k=1}^KT_km^3)=\mathcal{O}(nm^3)$ (linear in $n$) for computing the posterior for any given values of the parameters.

\subsection{Implementation using collapsed models}
\label{ImplCollNNGP}
The Bayesian hierarchical model in \eqref{eq:hierModel}, either with $\bC_{\btheta}$ or with $\tilde{\bC}_{\btheta}$ in the prior for $\bw$, allows full posterior inference for $\{\bbeta,\bw,\btheta,\tau^2\}$ using Markov chain Monte Carlo (MCMC). Gibbs sampling with random walk Metropolis steps provide full conditional distributions in closed form for $\{\bbeta,\bw\}$ and also for $\tau^2$ with an $\IG(a_{\tau},b_{\tau})$ prior. However, this convenience is nullified in practice by strong autocorrelation and poor mixing of the chains \citep{liu1994covariance}. %Such 
Samplers based on spatial DAG-based models have been devised, explored and compared in \cite{finley2019efficient}% and collapsed samplers have been seen to be considerably improve convergence
. 
Instead of (\ref{eq:hierModel}), we sample from
\begin{equation}\label{eq:hierModelCollapsed}
 p(\bbeta,\btheta,\tau^2\given \by) \propto p(\btheta,\tau^2)\times N(\bbeta\given \bmu_{\beta},\bV_{\beta})\times N(\by\given \bX\bbeta, \tilde{\bC}_{\btheta} + \tau^2\bI_n)\;,
\end{equation}
which is derived from (\ref{eq:hierModel}) by integrating out $\bw$, thereby ``collapsing'' the parameter space to a much smaller domain without $\bw$. This considerably improves mixing and convergence.

%In the sequel, we describe the implementation with collapsed likelihoods in the specific context of temporal processes. In particular, we describe some computational shortcuts linked to convenient patterns arising from the temporal structure. 

We will need to compute the inverse and determinant of $\tilde{\bLambda} = \tilde{\bC}_{\btheta} + \tau^2\bI_n$, which is $n\times n$. While $\tilde{\bLambda}^{-1}$ does not share the same convenient factorization of $\tilde{\bC}^{-1}$ and is also not guaranteed to be sparse, the Sherman-Woodbury-Morrison formulas reveal
\begin{equation}\label{eq:omegadef}
   \tilde{\bLambda}^{-1}=\tau^{-2}\IdentityMat-\tau^{-4}\bOm^{-1}, \quad \text{with} \quad \bOm = \tilde{\bC}^{-1}+\tau^{-2}\IdentityMat\;,
\end{equation}
where $\bOm$ enjoys the same sparsity as $\bC^{-1}$. 
Moreover, $\text{det}(\tilde{\bLambda})=\tau^{2n}\text{det}(\tilde{\bC})\text{det}(\bOm)$. The core of the algorithm is therefore to compute $\tilde{\bLambda}^{-1}$ through $\bOm$. In our application, the random effect is assumed to be the realization of $K$ independent temporal processes. As discussed in Section \ref{secindNNGP}, this implies a block-diagonal structure for $\tilde{\bC}$ that can be shown to be shared also by $\bOm$ (see Eq. \eqref{eq:omegadef}).
Each block $\bOm_k$ of $\bOm$ can be computed independently for each individual and the same holds for its inverse and its determinant. This means that the body of the algorithm will consist of a loop over all the individuals, which allows for straightforward parallelization. Unlike in spatial DAGs \citep{datta2016hierarchical, finley2019efficient}, we do not need fill-reducing permutation methods since neighbors sets for temporal processes consist of contiguous observations and $\left\lbrace\bOm_k\right\rbrace_{k=1}^K$ are banded matrices with no gaps.

%The considered algorithm is a \textit{Metropolis-withing-Gibbs}, where the variance and covariance parameters are updated through a random walk block-Metropolis step. In order to obtain a good acceptance rate ($\approx 25\%$) and convergence speed, the variance matrix of the random-walk Metropolis proposal should be appropriately tuned. However, the optimal tuning strongly depends on the considered data (size, number of covariates, number of parameters) and should be repeated in any single context and application. 
We devised a Gibbs sampler with Metropolis random walk updates for (\ref{eq:hierModelCollapsed}), where $\bbeta$ is updated from its full conditional distribution, while $\{\btheta,\tau^2\}$ are updated using an adaptive Metropolis step based on \cite{haario2001adaptive}. Here, after the first few iterations, a new proposal covariance matrix is regularly computed on the run according to the empirical covariance of the current chain. Subsequently, a mixture of the original and adaptive proposal is used as the new proposal. Convergence toward the desired acceptance rate is assured for an appropriate choice of the variance terms and of the adaptation rule \citep{roberts2009examples}. The algorithm has been coded using the \texttt{R 4.0.5} statistical environment. All expensive computations are managed by the \texttt{Eigen} library (version 3.3.7), which provides efficient routines for numerical linear algebra with an emphasis on sparse matrices. Our implementation of (\ref{eq:hierModelCollapsed}) outperforms the algorithms that update $\bw$ in terms of computational speed as it is implemented in the \texttt{spNNGP} package \citep{finley2017spnngp}. We present these comparisons in %Section~\ref{additionalsimulations} of 
the Supplementary Materials \citep{alaimo2023aoasSupplement} %\textcolor{purple}
{including a link to the GitHub repository hosting codes to implement the models}.

\subsection{Including spatial effects}
\label{sec:modSpatEff}
Accounting for spatial information in our Actigraph dataset presents some new considerations. As mentioned in Section~\ref{sec:intro}, spatial information is available to us in terms of the physical location along the trajectory as well as through covariates that are functions of space. Considering the discussion in Section~\ref{secMod}, the analytical goals of this dataset suggest accounting for spatial heterogeneity. Here, as argued earlier, modeling $\bw(\cdot)$ in (\ref{eq:tempMod1}) as a spatio-temporal process, including scalable versions, has challenges given that: (i) the trajectory's domain does not have a positive area; and (ii) associations among the measurements are more amenable to the temporal scale. Therefore, we introduce spatial effects into the mean employing a smooth function of space, $f_S(\cdot):\mathbb{R}^2\rightarrow \mathbb{R}$, approximated by a spline basis representation \citep[see, e.g.,][]{goodman2006refinable, ramsay2007applied}. For instance, if $J_x$ and $J_y$ are the dimensions of independently defined B-spline basis expansions on the $x$ and $y$ coordinates, respectively, then $f_S\left((x,y)\right)\approx\tilde{f}_S\left((x,y)\right)=\sum_{j_X=1}^{J_X}\sum_{j_Y=1}^{J_Y}\beta_{S,(j_X,j_Y)}B_{x,j_x}(x)B_{y,j_y}(y)$,
% \begin{equation*}
%     f_S\left((x,y)\right)\approx\tilde{f}_S\left((x,y)\right)=\sum_{j_X=1}^{J_X}\sum_{j_Y=1}^{J_Y}\beta_{S,(j_X,j_Y)}B_{S,j_X}(x)B_{S,j_Y}(y)
% \end{equation*}
where $B_{x,j_X}=\left[\bB_{x}\right]_{j_x}$ and $B_{y,j_Y}=\left[\bB_{y}\right]_{j_y}$ are the $j_x$-th and $j_y$-th element of the B-spline basis along the two axis. For any location $(x, y)\in\mathbb{R}^2$ the elements of the previous sum can be more compactly expressed through the tensor product basis $\bB_S(x, y)=\left(\bB_x\otimes\bB_y\right)(x, y)$. The size of this basis is $J_S=J_x\cdot J_y$ and depends on the size of the two original spline basis, which in turn depends on the chosen number of knots ${knots}_x, {knots}_y$ and degree ${deg}_x, {deg}_y$ (namely $J_c = {knots}_c+{deg}_c$ for $c=x,y$).
We now modify (\ref{eq:tempMod1}) to include the spline,
\begin{equation}\label{modSpec1}
    \bY(t) = \bX(t, \bgamma(t))\bbeta + \bB_S\left(\bgamma(t)\right)\bbeta_{S} + \bw(t) + \beps(t), \qquad t\in\mathcal{\mathbb{R}^+}\;,
\end{equation}
where $\bgamma(t)=\left\{\bgamma_1(t),\bgamma_2(t),\ldots,\bgamma_K(t)\right\}$, $\bgamma_k(t) = (\gamma_{k,x}(t),\gamma_{k,y}(t)):\mathbb{R}^+\rightarrow \mathbb{R}^2$ is the trajectory function mapping time $t$ for individual $k$ to its position and $\bB_S\left(\bgamma(t)\right)$ is the $K\times J_S$ matrix with row $k$ corresponding to the $J_S$ basis elements for the coordinates at time point $t$ for individual $k$. A proper choice of $J_S$ (i.e. knots and degree) is required to fit a spline surface flexible enough to describe the spatial variations at the scale of interest without incurring over-fitting.
Let $\bB=\bB_S\left(\bgamma(\calT)\right)$ be the $n\times J_S$ matrix containing the B-spline basis elements evaluated at the observed location of each individual $\bgamma(\calT)=\left\{\gamma_1(t_{11}), \gamma_1(t_{12}),\dots,\gamma_K(t_{KT_k})\right\}$. Following Equation \eqref{eq:hierModelCollapsed}, we sample from %the posterior,
\begin{equation}\label{eq:hierModelCollapsedShrink}
 p(\bbeta,\bbeta_S,\btheta,\tau^2\given \by) \propto p(\btheta,\tau^2)\times p_S(\bbeta_S)\times N(\bbeta\given \bmu_{\beta},\bV_{\beta})\times N(\by\given \bX\bbeta+\bB\bbeta_S, \tilde{\bC}_{\btheta} + \tau^2\bI_n)\;,
\end{equation}
where the prior $p_S(\cdot)$ needs to be specified. The Actigraph data includes millions of observations in a limited study area, of which some assume different values in the same location (or in its immediate vicinity) so over-fitting will not be an issue. However, some areas present sparsely observed points (trajectories are not uniformly distributed, as shown in Figure \ref{ObsWW}). This may cause coefficients corresponding to those regions to be weakly identified. To control for the balance of all these components, we may assign ad-hoc priors to the spatial spline regression coefficients \citep{eilers1996flexible} for penalizing deviation from a certain degree of smoothness and favoring identifiability. This behavior suggests the Bayesian P-Spline \citep{hastie2000bayesian, lang2004bayesian}. While keeping the Gaussian priors, we effectuate shrinkage by choosing a suitable precision matrix $\bP$ and introducing a shrinkage parameter $\lambda$ at a deeper level of the hierarchy. To be precise, $\displaystyle \bbeta_S \given\lambda\propto \exp\left\lbrace-\frac{\lambda}{2}\cdot\bbeta_S\bP\bbeta_S^\top\right\rbrace$ and $\lambda\sim \mathcal{G}(\alpha_\lambda, \beta_\lambda)$.
We consider two possible forms for $\bP$, which imply different penalization for the coefficients:
\begin{itemize}
    \item \textbf{Ridge-like} prior, which is to say $\bP=\bP_{RL}=\IdentityMat_{J_S}$;
    \item \textbf{First-order random walk} prior, which is to say:
    \begin{equation*}
        \bP=\bP_{RW}:\; [\bP_{RW}]_{ij}=\begin{cases}
            n_i\quad &i=j\\
            -1\quad &i\sim j\\
            0\quad &\text{otherwise}
        \end{cases}
    \end{equation*}
    where $n_i$ is the number of neighbors of knot $i$ and $i\sim j$ denotes a neighboring relationship between the knots.
\end{itemize}
Both precision matrices provide a multivariate Gaussian prior distribution on the coefficients. However, the latter is improper since $\text{rank}\left(\bP_{RW}\right)<J_S$. Nevertheless, if we collect the B-Spline basis elements with the other covariates as $\bX^*=\left[\bX, \bB\right]$ and stack the corresponding coefficients into the joint vector $\bpsi=\left[\bbeta, \bbeta_S\right]$, then the posterior distribution of the latter is a proper multivariate Gaussian with full conditional distribution $\bpsi\given \cdot \propto \Norm_J\left(\bpsi\left|\bG^{-1}\bg,\, \bG^{-1}\right.\right)$, where $\bG = {\bX^*}^\top \tilde{\bLambda}^{-1} {\bX^*} + \bV_{\bpsi}^{-1}$ and $\bg =  {\bX^*}^\top \tilde{\bLambda}^{-1} \by + \bV_{\bpsi}^{-1} \bmu_{\bpsi}$ with $\bV_{\bpsi}^{-1}=\text{diag}\left(\bV_{\bbeta}^{-1},\, \lambda\cdot\bP\right)$ and $\bg = \left[\bmu_{\bbeta},\,\bmu_{\bbeta_S}\right]^\top=\bzero^\top$.
Moreover, the Gamma prior on $\lambda$ implies a Gamma full-conditional distribution $\displaystyle \lambda\given \cdot \propto \mathcal{G}\left(\lambda\left|\alpha_\lambda+1/2,\,\beta_\lambda+\bbeta_S^\top\bP\bbeta_S\right.\right)$. 

To estimating the model in \eqref{modSpec1}, we jointly update $\bpsi$ and $\lambda$ from their full conditional distributions. In particular, the Gibbs' sampling step can be adapted to get full inference also on the spline coefficients $\bbeta_S$ and the shrinkage parameter $\lambda$ \citep[see the Supplementary Material][]{alaimo2023aoasSupplement}. In practical terms, this requires $J_S$ additional linear coefficients to be estimated, whose size $p^{*}=p+J_S$ may undermine the efficiency of the algorithm. For example, calculations in Step~1b are quadratic w.r.t. $p^{*} \rightarrow \mathcal{O}(n{p^{*}}^2)$. %Hence, we must define efficient strategies to limit the additional computational burden.
Steps~1a~and~1b (i.e. the most expensive in $p^{*}$) are executed in the first iteration and subsequently, only in those iterations where new values of $\btheta$ are accepted. When $\btheta$ is rejected, we retain in memory the previously computed value (which would stay unchanged). Thus, if we attain an optimal acceptance rate of $\approx 20\%-30\%$ in the Metropolis Hastings step on $\btheta$, the computation is avoided in the majority of cases with a sensible improvement in computation time and speed.

\subsection{Simulations} \label{secExp}
We conducted simulation experiments to evaluate the model described in Section~\ref{sec:modSpatEff} and compared the performance of our algorithm in terms of fitting, prediction error and computational speed with other routines available from the \texttt{spNNGP} package \citep{finley2017spnngp}. Additional comparative experiments are provided in the Supplementary Material \citep{alaimo2023aoasSupplement}. We executed our MCMC algorithms on %Cluster Terastat, an HPC infrastructure developed by the Department of Statistical Sciences (DSS) in the University of Rome \textit{``La Sapienza''}, in collaboration with CINECA, for the resolution of mathematical and statistical models on big data. It is 
a computing environment
%currently 
equipped with 12 modern computational nodes with 32 cores each, roughly equivalent to 3 TeraFlop/sec, and $256$Gb of RAM% (\url{https://www.dss.uniroma1.it/en/node/5870/technical-specifications})
. Each of the presented applications have been executed on a single node exploiting the computational power of all cores. The results presented are based upon posterior samples that were retained after diagnosing convergence using visual tools (e.g., traceplots, autocorrelation), effective sample sizes, Monte Carlo standard errors (MCSE) and other diagnostics offered by the \texttt{coda}, \texttt{mcse} and \texttt{bayesplot} packages in the \texttt{R} computing environment; the Supplementary Material \citep{alaimo2023aoasSupplement} includes specific details. 

\begin{figure}[ht]
    \centering
    \begin{subfigure}[b]{0.42\linewidth}
      \includegraphics[width=0.9\textwidth]{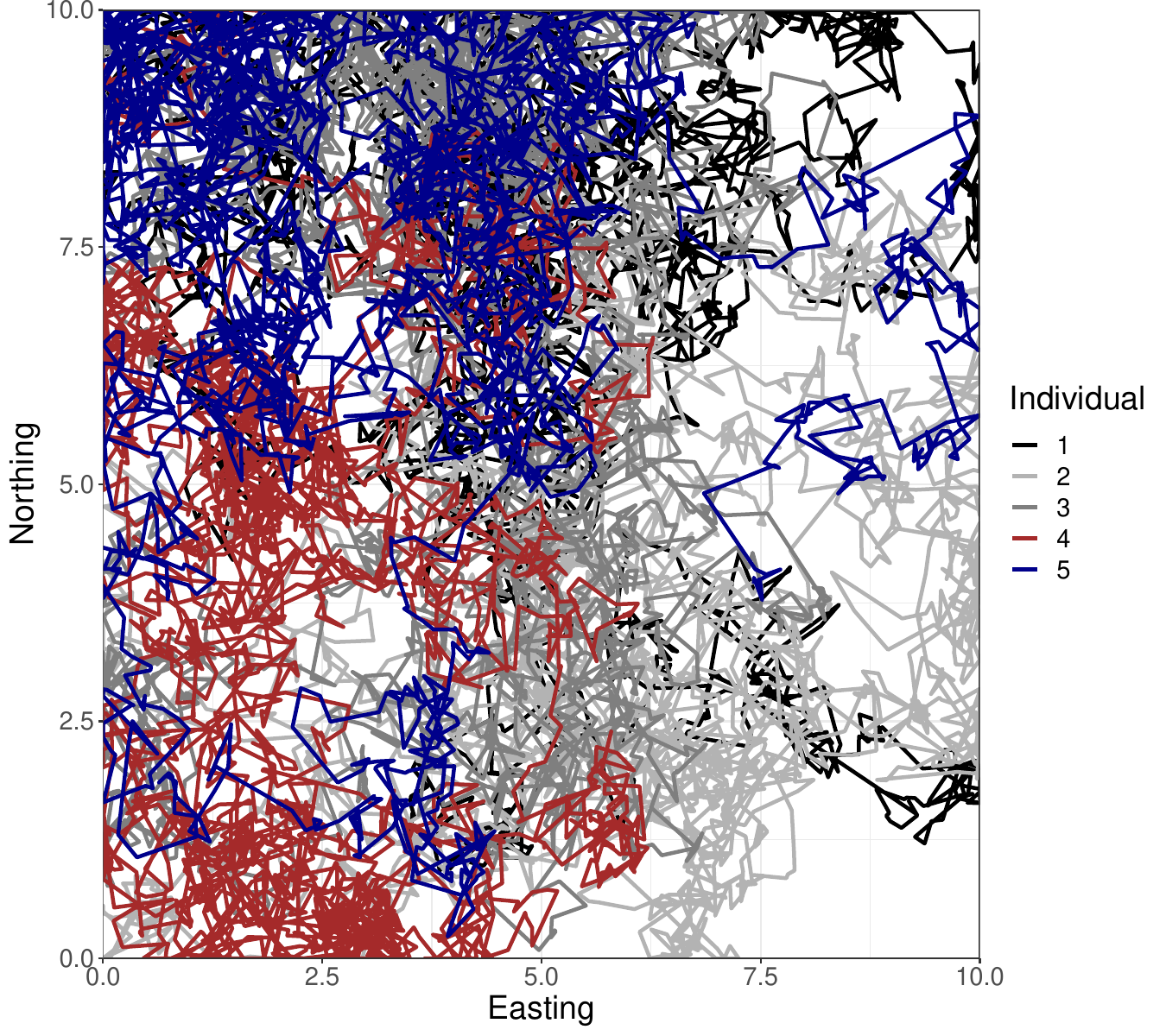}
    \caption{Trajectories}
    \label{fig:IndTrajSim}
    \end{subfigure}
    \hspace{0.1cm}
    \begin{subfigure}[b]{0.42\linewidth}
      \includegraphics[width=0.9\textwidth]{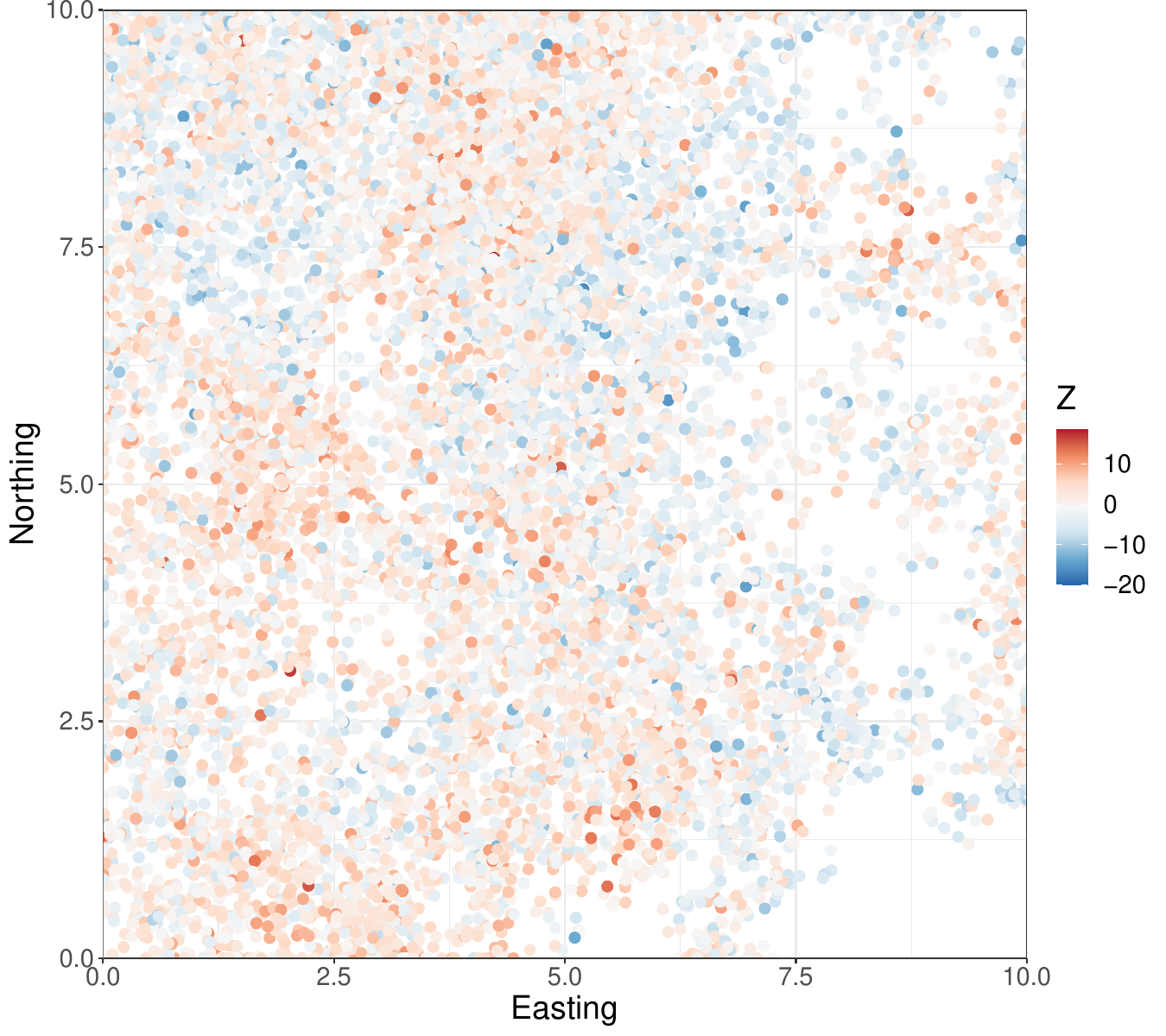}
    \caption{Points}
    \label{fig:IndPointsSim}
    \end{subfigure}
    \caption{Observed trajectories (a) and observed points (b) for the simulated dataset.}
    \label{SimIndTrajPointsEExp3}
\end{figure}

%We demonsrate our model's ability to disentangle the true temporal and spatial components, recovering the parameters of the temporal covariance and the spatial effect over the region of interest, when observing points belonging to random trajectories evolving in space. 
We first generated $T_k=2\times 10^5$ time points for $K=5$ individuals, where each time point $t_{ki}$ followed exponential waiting times between observations, i.e. $t_{ki} = \sum_{h=1}^{i-1} \delta_{h}$, and $\delta_{h} \stackrel{iid}{\sim} \mbox{Exp}(5)$. Given the time points, we constructed spatial trajectories $\bgamma_k(\cdot)$, $k=1,\ldots,K$, by simulating $\bs_k=\left[\gamma_k(t_{k1}),\dots,\gamma_k(t_{kT_k})\right]^{\top}$, where subsequent components were independent Gaussian random walks over the square $\mathcal{S}=(1,10)\times (1,10)$, with the variance of each step along the horizontal and vertical axis proportional to the elapsed time between two subsequent observations. If the trajectory left the square, it was projected onto the border and the next step would resume from there. The simulated trajectories are shown in Figure~\ref{fig:IndTrajSim}.

\begin{table}[t]
    \centering
    {\fontsize{8}{8}\selectfont
    \begin{tabular}{lccccccc}
    \toprule
        \multirow{2}{*}{\textbf{Param. (True)}} & \multicolumn{2}{c}{\textbf{S-Spline}} & \multicolumn{2}{c}{\textbf{P-Spline}}\\
         &  Point & Interval & Point & Interval \\
        \midrule
         $\beta_{01}\; (-3.76)$ & -3.799 &  (-3.846,-3.752) & -3.797 &  (-3.844,-3.75) \\
         $\beta_{02}\; (0.65)$  &  0.572 &  (0.523,0.62)   & 0.575 &  (0.526,0.623)   \\
         $\beta_{03}\; (-0.60)$ & -0.649 &  (-0.697,-0.6) & -0.646 &  (-0.693,-0.598)  \\
         $\beta_{04}\; (2.36)$  & 2.326  &  (2.277,2.374)   & 2.328  &  (2.28,2.376)   \\
         $\beta_{05}\; (-0.33)$ & -0.359 &  (-0.408,-0.31) & -0.356 &  (-0.404,-0.308) \\
         $\beta_1\; (2.59)$  &  2.599 &  (2.59,2.608)   & 2.599  &  (2.59,2.608)  \\
         $\beta_2\; (2.70)$ & 2.691 &  (2.683,2.7) & 2.691 &  (2.683,2.7)  \\
         $\beta_3\; (-0.58)$  & -0.586  & (-0.595,-0.577)   & -0.586  &  (-0.595,-0.577) \\
         $\sigma^2\; (1)$    & 1.001 & (0.973,1.032)   & 0.993 &  (0.965,1.023) \\
         $\phi\; (1)$        & 0.994 & (0.948,1.04)   & 1.01  &  (0.964,1.063) \\
         $\tau^2\; (1)$      & 1.001 & (0.984,1.018)   & 1.001  &  (0.984,1.018)\\
         \midrule
         \textbf{Metric} & Out-of-sample & In-sample & Out-of-sample & In-sample \\
         \midrule 
         Coverage    &  0.95        & 0.99        &  0.95        &  0.99\\
         RMSPE (r)   &  0.07 (1.18) & 0.03 (0.84) &  0.07 (1.19) & 0.03 (0.84)\\
         PIW         &  4.66        & 4.44        &  4.66        & 4.44 \\
         DIC         &  \multicolumn{2}{c}{115'543} & \multicolumn{2}{c}{115'556} \\
         \midrule 
         Fitting time (h) & \multicolumn{2}{c}{2.18}   & \multicolumn{2}{c}{ 2.2} \\
         \bottomrule
    \end{tabular}
    }
    \caption{Parameter estimates, predictive validation and fitting times (hours) on the simulated dataset for all the considered models.}
    \label{tab:ResTableExp3}
\end{table}

Given the time points and positions (Figure~\ref{fig:IndPointsSim}), we generated the latent temporal Gaussian processes $w_k(\cdot)\stackrel{ind}{\sim} \mathcal{GP}(0,c_{\btheta}(\cdot,\cdot))$ with an exponential covariance $\displaystyle c_{\btheta}(t,\, t') = \sigma^2 \exp\{-\phi\cdot |t-t'|\}$, where $\sigma^2 > 0$ represents the variance of the process, $\phi > 0$ is the decay in temporal correlation (range) and $\tau^2 > 0$ the residual variance (nugget). The spatial effects are then introduced through $f_S(\cdot):\mathcal{S}\rightarrow \mathbb{R}$ by considering a tensor product spline basis of degree $2$ and with $9$ knots over the square domain (including boundary knots), where the spline coefficients $\bbeta_S$ have been fixed to randomly generated values from $\mathcal{N}_{81}\left(\bzero,\lambda\IdentityMat_{81}\right)$ with $\lambda=0.5$.
The model also included individual-specific intercepts $\left\lbrace\beta_{0k}\right\rbrace_{k=1}^5$ and the effect of 3 covariates with random values drawn independently at each location from a $\mathcal{N}(0,1)$ distribution, leading to covariate vectors $\left\lbrace\bx_{ki}\right\rbrace_{i=1}^{T_k},\; k=1,\dots, K$. 
The effect of the covariates is assumed common across individuals, and set to be determined by slopes $\bbeta=[\beta_1, \beta_2, \beta_3]^{\top} $.

We generated values of the outcome for individual $k$ at time $t_{ki}$ and location $\bs_{ki}=\gamma_k(t_{ki})$ according to the generative process defined by \eqref{modSpec1} with parameters fixed as above. This yielded
% from $y_{ki} = \beta_{0k} + \bx_{ki}^{\top}\bbeta + f_S(\bs_{ki}) + w_k(t_{ki}) + \epsilon_{ki}$ with $\epsilon_{ki}\overset{iid}{\sim}\mathcal{N}\left(0,\tau^2\right)$ for each $i$ and $k$. 
a simulated dataset $D_{sim}=\left\lbrace(\text{Ind}_j, t_j, \bs_j, y_j, \bx_j^\top)\right\rbrace_{j=1}^n$ with $n=10^5$ observations, where $\text{Ind}_j$ denotes the individual corresponding to row $j$. Then, we fit the model in \eqref{eq:hierModelCollapsedShrink} on $70\%$ of the total observations in $D_{sim}$.
The remaining $30\%$ were held out to assess out-of-sample predictive performances in terms of %\textit{Relative Mean Squared Prediction Error} (RMSPE), 
\textit{Relative and Root Mean Squared Prediction Error} (RMSPE), \textit{Coverage}, and \textit{Predictive Interval Width} (PIW).
Intercept and slope regression parameters were assigned $\Norm(0,\, 10^6)$ priors; the variance components, $\sigma^2$ and $\tau^2$, were both assigned inverse Gamma $\IG(2,2)$ priors; and the decay parameter $\phi$ received a Gamma prior $\mathcal{G}(1,1)$. For the spline coefficients, we considered both the penalized versions in Section \ref{sec:modSpatEff}. The first is referred to as an S-Spline (shrinking splines), and the second as P-Spline (penalized splines).

\begin{figure}[ht]
    \centering
    \begin{subfigure}[b]{0.45\linewidth}
    \centering
    \includegraphics[width=.9\textwidth]{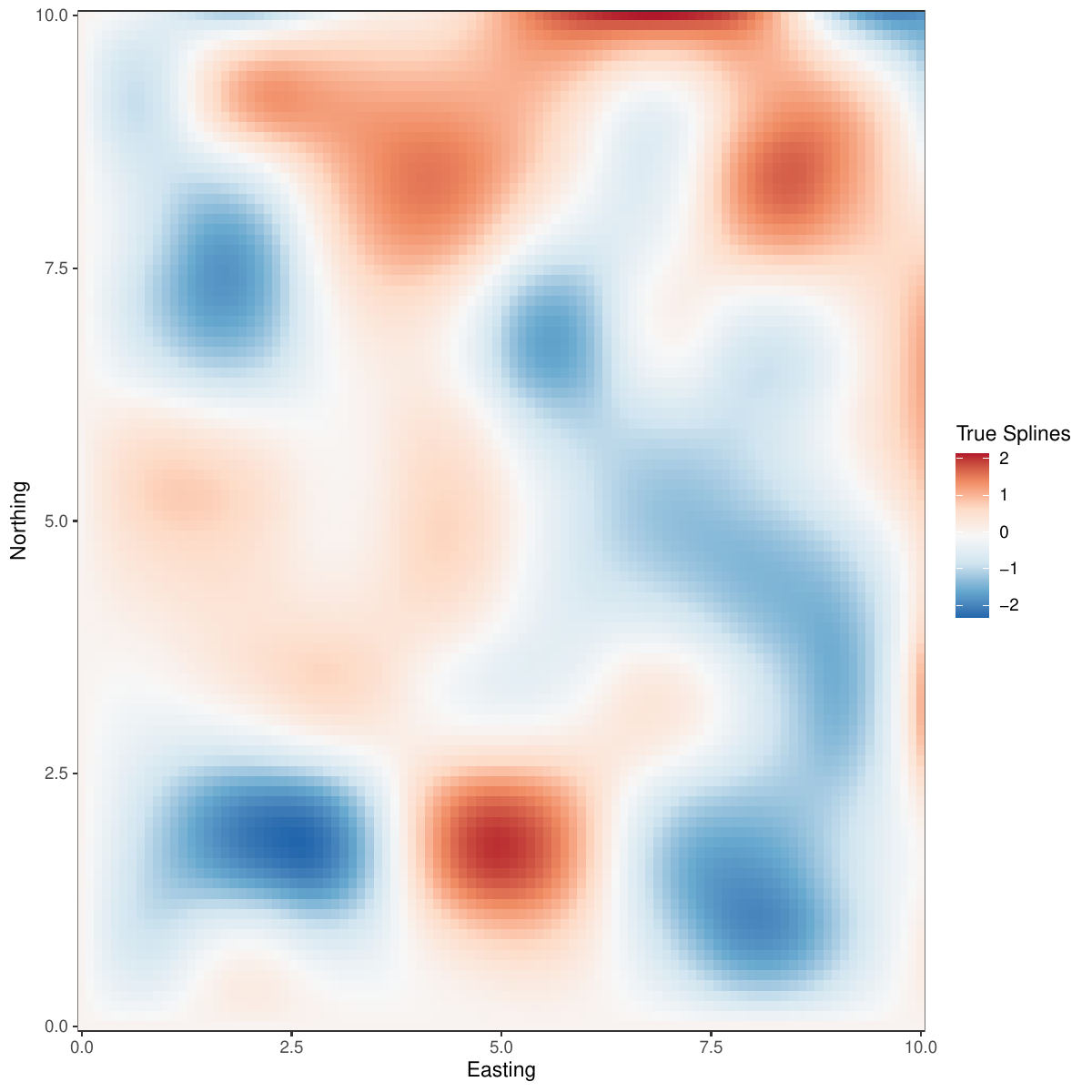}
    \caption{True splines surface}
    \label{fig:TrueSimSplines}
    \end{subfigure}
    \hspace{0.15cm}
    \begin{subfigure}[b]{0.45\linewidth}
    \centering
    \includegraphics[width=.9\textwidth]{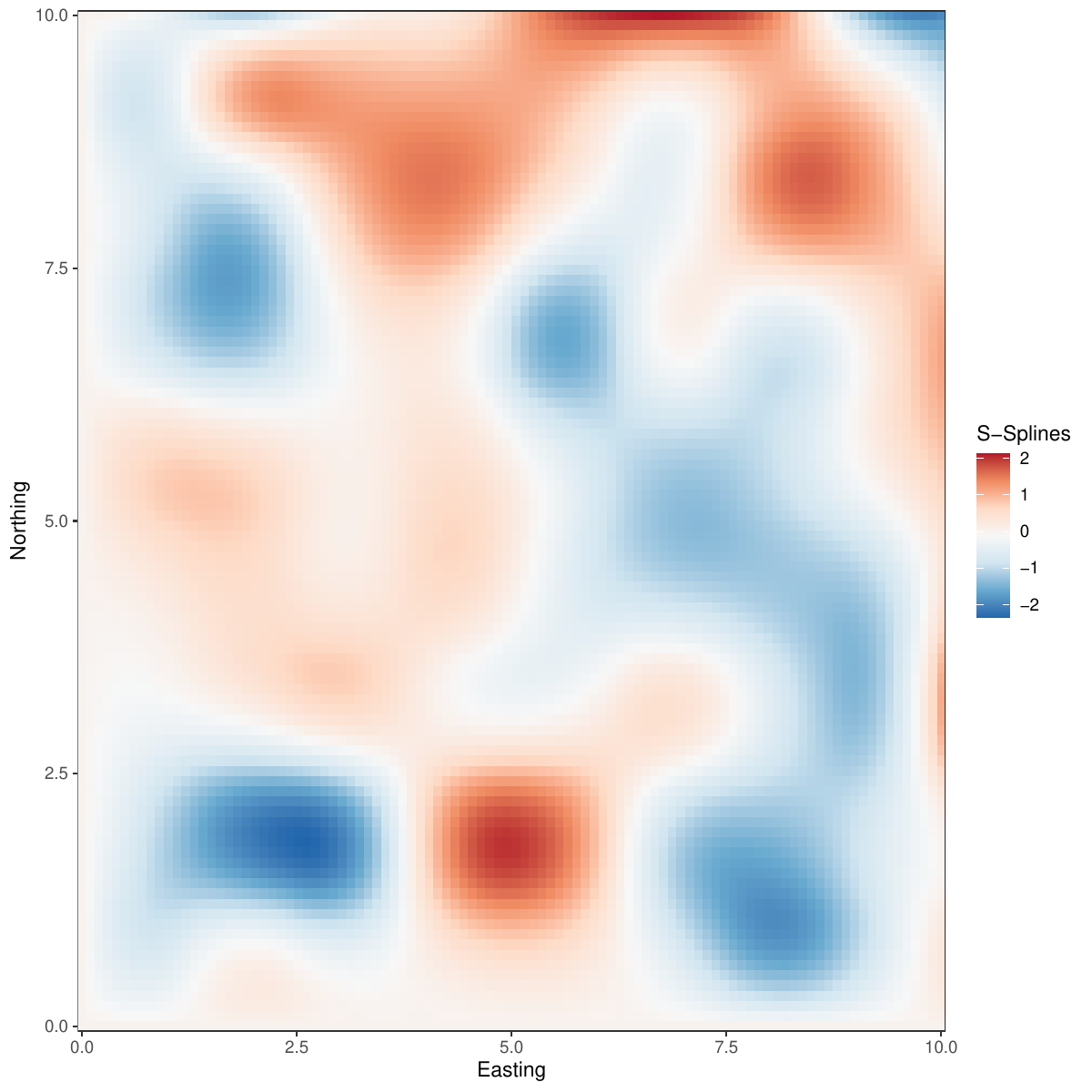}
    \caption{Estimated S-Splines}
    \label{fig:EstSSplines}
    \end{subfigure}
    % \hspace{0.15cm}
    % \begin{subfigure}[b]{0.3\linewidth}
    % \centering
    % \includegraphics[width=\linewidth]{images/EstimatedPSplinesSurface.pdf}
    % \caption{Estimated P-Splines}
    % \label{fig:EstPSplines}
    % \end{subfigure}
    \caption{True and estimated spline surfaces using S-Splines.}
        \label{fig:SSSurfaces}
\end{figure}

Table~\ref{tab:ResTableExp3} presents the posterior estimates. We also included the Deviance Information Criterion (DIC) for both models. Performances in the two settings are almost identical, but the DIC favors the S-Spline model. This is not surprising as the data were generated using an analogous shrinkage prior for the $\bbeta_S$'s. Further details, including the estimates of the spline coefficients are provided in the Supplementary Material \citep{alaimo2023aoasSupplement}. %\textcolor{purple}
{Figure~\ref{fig:SSSurfaces} presents the posterior estimate of the spatial surface. We compare the true latent surface with just the S-Splines as it performs slightly better with respect to the DIC, but notice that P-Splines provides practically identical estimates.}
%Figure~\ref{fig:DiffSS} displays the differences between the true spline surface and the estimates; very negligible differences (a small upward bias) are seen. 

% \begin{figure}[ht]
%     \centering
%     \begin{subfigure}[b]{0.4\linewidth}
%     \centering
%     \includegraphics[width=\linewidth]{images/BoxplotSplineDifferenceSS.pdf}
%     \caption{Difference distribution}
%     \label{fig:BoxDiffSSplines}
%     \end{subfigure}
%     \hspace{0.15cm}
%     \begin{subfigure}[b]{0.4\linewidth}
%     \centering
%     \includegraphics[width=\linewidth]{images/SSplinesSurfaceDifference.pdf}
%     \caption{Spatial distribution of error}
%     \label{fig:DiffPSplines}
%     \end{subfigure}
%     \caption{}
%     \label{fig:DiffSS}
% \end{figure}

\section{Application}\label{sec:App}
%\textcolor{purple}
{We apply the proposed model in (\ref{modSpec1}) to estimate the MAG (measured in G) for participants in the study accounting for subject-specific features and spatial effects on the mean, while modeling the latent temporal dependence as described in Section~\ref{sec:tempMod}. We split the data into training ($70\%$) and testing ($30\%$) subsets, where the records have been allocated to each subset according to a random sample stratified by individual. The testing set is used to assess the out-of-sample predictive performances in terms of \textit{Relative Mean Squared Prediction Error} (RMSPE), \textit{root Mean Squared Prediction Error} (rMSPE), \textit{Coverage} (Cov), and \textit{Predictive Interval Width} (PIW). Posterior inferences are based on $5,000$ samples retained after diagnosed convergence from $10,000$ MCMC iterations.}

%We analyze activity levels throughout the ``active time'' -- when the Actigraph device records the individual as being physically active -- excluding epochs when the device was not worn or when there was no physical activity (e.g., the individual was sitting or lying down). 
% The data processing and merging of actigraph data with GPS locations resulted in two final datasets (Section~\ref{secData}). These are treated separately. In both applications, 
%$70\%$ of the total observations are used for training the model, while the remaining are excluded to assess the out-of-sample predictive performances in terms of \textit{Relative Mean Squared Prediction Error} (RMSPE), \textit{Root Mean Squared Prediction Error} (rMSPE), \textit{Coverage}, \textit{Predictive Interval Width} (PIW).

\subsection{Model specification}

% \begin{comment}
% \begin{figure}[t]
%     \centering
%     \begin{subfigure}[b]{0.45\linewidth}
%     \centering
%     \includegraphics[width=0.8\linewidth]{images/OrigMAG.pdf}
%     \caption{MAG}
%     \label{plotMAGTemp}
%     \end{subfigure}
%     \begin{subfigure}[b]{0.45\linewidth}
%     \centering
%     \includegraphics[width=0.8\linewidth]{images/LogMAG.pdf}
%     \caption{log(MAG)}
%     \label{plotlMAGTemp}
%     \end{subfigure}
%     \caption{Observed $MAG$ (a) and $lMAG$ (b) in the whole sample.}
% \end{figure}
% \end{comment}

Spatial effects are introduced by considering the tensor product of two analogous univariate B-spline basis on %longitude and latitude
each spatial axis. %
%\textcolor{purple}
{After a preliminary validation through the DIC,} we choose two bases of degree $3$ with $12$ equally spaced knots over a square encompassing Westwood. This sums up to $J_S= (9+3)\times (9+3) = 144$ terms for our complete spline basis, including the boundary knots. 
\begin{comment}
\textcolor{purple}{The full process specification for individual $k=1,\dots, K$ is:
\begin{equation}
    \bY(t) = \bX\lrnd t, \bgamma(t)\rrnd\bbeta + \sum_{j=1}^{J_S}\beta_{S,j}B_{S,j}\left(\bgamma(t)\right) + \bw(t) + \beps(t), \qquad t\in\mathcal{\mathbb{R}^+},
\label{modSpecFull}
\end{equation}
where:
\begin{itemize}
    \item locations are described as functions of time through the individual trajectory functions $\gamma_k(\cdot),\, k=1,\dots,K$ and $\bgamma(\cdot)=\lsq \gamma_1(\cdot),\dots,\gamma_K(\cdot)\rsq$;
    \item $\bX(\cdot, \cdot)$ collects all the individual and spatial covariates, potentially varying in time and/or space;
    \item $B_S=B_X\otimes B_Y$ is the tensor product bivariate spline;
    \item $\bw(t)$ is assumed to by a zero-mean stationary Gaussian process $\GP\lrnd c_\theta\lrnd\cdot,\cdot\rrnd\rrnd$ with an exponential covariance function.
\end{itemize}
}
\end{comment}
%\textcolor{purple}
{All numerical variables in $\bX(\cdot,\cdot)$ have been standardized for improving the efficiency of the MCMC sampling \citep{gilks1996strategies}.}
The presence of temporal dependence in individual trajectories was investigated through an individual-specific exploratory analysis on the residuals from a standard linear regression and an {Ornstein-Uhlenbeck process} (GP with an exponential covariance function) was specified to capture temporal dependence as a parsimonious and effective model for the behavior of the underlying residual process. 

Finally, the outcome is log-transformed %to $lMAG_k(t)=\log(\text{MAG}_k(t))$ for $k=1,2,\ldots,K$ and $t=t_{k1},\ldots,t_{kT_k}$ 
in order to comply with the Gaussianity assumption of the model.
% The original covariates have been enriched with a binary variable indicating if the measures refer to the period before or after a Bruin Bike Share (BBS) program was launched in Westwood, Los Angeles, to account for the effect of a new specific policy which aims at improving the physical activity level of the participants.
% \begin{equation}
% lMAG_k(t)=\log(MAG_k(t)),\qquad k=1,\dots,K,\quad t=t_{k1},\dots,t_{kT_k}
% \end{equation}
%\vspace{0.5cm}
%\textcolor{purple}{TWO ALTERNATIVES HERE}
%\textcolor{purple}
{
%\begin{itemize}
    %\item OLD VERSION UPDATED. 
    We denote the parameter associated with variable ``$\text{varname}$'' as $\beta_{\text{varname}}$ and the levels of each categorical covariate as $\text{varname}_{(j)}$ for $j=1,\dots, J_{\text{varname}}$. Hence,
\begin{equation}
    \label{eq:TempMod}
    \begin{split}
    \mathbb{E}\lsq \log(MAG_k(t))\rsq &= \beta_0 + \sum_{j=2}^{J_{\text{BMI}}}\beta_{\text{BMI},j}\cdot\mathbb{I}\left(\text{BMI}_k=\text{BMI}_{(j)}\right) + \sum_{j=2}^{J_{\text{Sex}}}\beta_{\text{Sex},j}\cdot\mathbb{I}\left(\text{Sex}_k=\text{Sex}_{(j)}\right)+\\ 
    &\quad +\sum_{j=2}^{J_{\text{Age}}}\beta_{\text{Age},j}\cdot\mathbb{I}\left(\text{Age}_k=\text{Age}_{(j)}\right)+\sum_{j=2}^{J_{\text{Eth}}}\beta_{\text{Eth},j}\cdot\mathbb{I}\left(\text{Eth}_k=\text{Eth}_{(j)}\right)+\\ 
    &\quad +\beta_{\text{distHome}}\cdot\text{distHome}_k(\gamma_k(t))+\beta_{\text{NDVI}}\cdot\text{NDVI}(\gamma_k(t))+\\
    &\quad +\beta_{\text{distParks}}\cdot\text{distParks}(\gamma_k(t))+\beta_{\text{Slope}}\cdot\text{Slope}(\gamma_k(t))+\\
    &\quad +\sum_{j=1}^{J_S}\beta_{S,j}B_{S,j}\left(\gamma_k(t)\right)%\\
    \end{split}
    %&w_k(t)\overset{\text{ind}}{\sim}\NNGP\left(0, c_\btheta(\cdot,\cdot)\right)\\
    %&\epsilon_k(t)\overset{\text{ind}}{\sim}\mathcal{N}\left(0, \tau^2\right)
\end{equation}
where $\mathbb{I}(\cdot)$ denotes the indicator function, $w_k(\cdot)$ is the DAG-based approximation (Section~\ref{secindNNGP}) for $\mathcal{GP}(0,c_{\btheta}(\cdot,\cdot))$, and $\epsilon_k(t)\overset{\text{iid}}{\sim}\mathcal{N}\left(0, \tau^2\right)$. 
%\item ALTERNATIVE. Among the individual covariates we include the \textit{sex} of the individual, the \textit{Ethnicity}, the \textit{age class}, and the \textit{BMI} class. We use all the available spatial covariate, i.e. the individual specific \textit{distance from home}, and the shared \textit{Overlay Weighted Distance to Parks},\textit{NDVI}, and \textit{Slope}.   
%\end{itemize}
}
%\textcolor{purple}
{The baseline subject represents an underweight Asian female less than 18 years of age. Other socioeconomic factors (e.g. education and income level) have been excluded from the analysis as they are strongly associated with ethnicity and age% and resulted non-significant in some preliminary runs
.}
% \begin{figure}[t]
%     \centering
%     \begin{subfigure}[b]{0.42\linewidth}
%       \includegraphics[width=0.9\textwidth]{images/VariogPID109.pdf}
%     \caption{Individual $1$}
%     \label{fig:vario1}
%     \end{subfigure}
%     \hspace{0.1cm}
%     \begin{subfigure}[b]{0.42\linewidth}
%       \includegraphics[width=0.9\textwidth]{images/VariogPID139.pdf}
%     \caption{Individual $2$}
%     \label{fig:vario2}
%     \end{subfigure}
%     \caption{Variograms of the standard linear regression residuals on individuals $109$ and $139$.}
%     \label{fig:variograms}
% \end{figure}

\subsection{Prior distributions}
The prior choices for each set of parameters and/or coefficients followed ad-hoc strategies. We incorporated priors such as $\bbeta\sim\mathcal{N}_J\left(\bzero, 10^6\cdot\IdentityMat_J\right)$, $\sigma^2\sim \mathcal{IG}(2,2)$ and $\tau^2\sim \mathcal{IG}(2, 2)$ with $J$ being the total number of $\bbeta$ coefficients.  %\textcolor{purple}
{The high spatial density of observations in several areas of the map enables robust estimation of the spatial effects. However, over-fitting may emerge from the the high dimension of the spline basis. Furthermore, there are areas in Westwood that present sparsely observed data-points and the model could struggle to identify the spline coefficients referred to those areas and jeopardize convergence of the MCMC algorithm.}
%Given the reduced number of knots and the high spatial density of observations in several areas of the map% (see Figure~\ref{fig:obslocWW})
%, over-fitting is not a concern. However, there are also areas in Westwood that present sparsely observed data-points. The model can struggle to identify the spline coefficients referred to those areas and jeopardize convergence of the MCMC algorithm.
Hence, we consider the S-Spline (Ridge-like prior) described in Section \ref{sec:modSpatEff} to mitigate these potential issues, where the shrinkage parameter $\lambda$ has been assigned a $\mathcal{G}(1,1)$ prior.

\subsection{Results}\label{subsec: results_spatiotemporal}
%\textcolor{purple}
{Fitting the model required $\approx 7$ hours on a computer equipped with 2 processors AMD EPYC 7452, each one having 32 cores for a maximum of 256Gb of RAM. The acceptance rate obtained is $\approx 28\%$, supporting the consistency of our adaptive strategy.}
Table~\ref{tab:ResTableSpApp} presents parameter estimates and %\textcolor{purple}
{performance metrics} for the model %\textcolor{purple}
{in (\ref{modSpec1}) with the explanatory variables specified in (\ref{eq:TempMod})} alongside estimates from a %\textcolor{purple}
{Bayesian linear regression model that includes the spatial spline terms, but neglects the temporal dependence structure}.
% Conclusions on the regression coefficients are very similar to those from Section~\ref{sec:tempApp}. However, accounting for the spatial effects allow for easier interpretation of the age-group regression coefficients: 
%The Age-class coefficients show how the older the person and the lower is the expected physical activity level. Also, both models estimate the effect of BBS as trending slightly negative. This somewhat surprising finding can be attributed to a few factors. First, the observations after the BBS launch are mostly from the winter season (February to April, the coldest months in L.A. together with December), while the others include summer and autumn (June to November, the warmest months). Given that physical activity levels tend to be lower in the colder months, there is indication of some possible confounding between the BBS effect and seasonality. Second, not all subjects were exposed to the BBS after its launch and, hence, could not take advantage of it. Third, questions exist about the ability of a wrist-worn accelerometer to detect the activity exerted while biking.%A more thorough analysis on the effects of the BBS should require further data collection.

%In this application, the intercept is estimated to be $\approx 5.31$. This corresponds to a $MAG$ per minute count of $1214$. According to currently available literature, this would correspond to a \textit{moderate-to-vigorous} physical activity level, highlighting how the studied population is mostly composed of young and active individuals.

%\textcolor{purple}
{The estimates from the two models are largely consistent with each other, although accounting for temporal dependence tends to somewhat mitigate the effects of some predictors. We anticipate the temporal process to absorb the impact of certain predictors --- especially when their relationship with the dependent variable is complex and nonlinear --- and this appears to be the case with ``Slope'', which loses its significant positive impact on MAG once the temporal process is incorporated. All other variables seem to retain the nature of their impact on MAG. These coefficients are interpreted with respect to the baseline measure of an underweight 18 year old Asian female.}

%\textcolor{purple}
{The intercept represents the natural logarithm of the MAG for the baseline subject and reckons with both active as well as inactive time points for the high-resolution digitally filtered data. Since we model continuously over time, the resulting value of the MAG, which is about $\exp(-2.9)\approx 0.06$, corresponds to the low region of moderate physical activity level \citep[see the Supplementary Material,][]{alaimo2023aoasSupplement}. Other gleanings from Table~\ref{tab:ResTableSpApp} indicate that MAGs vary by ethnicity in the study cohort as Whites tend to record lower MAGs, while Latin Americans and African Americans tend to register larger MAGs compared to Asians. Impact of Age groups on MAGs also tend to vary with the groups of $25$-$34$ and $45$-$70$ year old subjects tending to register lower MAGs than the baseline (young), while the middle-aged group tends to be higher than the baseline. This is not entirely surprising because subjects in the $25$-$34$ year old group tend to exercise less than the younger (baseline) and middle-aged groups with $25$-$34$ year old subjects having less time as they embark on their careers (less free time), while those in the $45$-$70$ range also tend to follow a less vigorous lifestyle regimen due to their age. The effect of Body Mass Index (BMI) is also seen to vary based upon the categories of weight. While all three categories indicate a significantly higher MAG compared to the baseline, the impact of the overweight, but not obese, category seems higher than the other two. We do not fully know the extent to which larger body weights affect accelerometer readings, but this variation must also account for the fact that higher BMI may also correspond to muscular (not unfit) individuals engaging in more vigorous lifestyle regiments. The spatially-indexed predictors indicated the expected positive impact of NDVI (more greenness encourages more outdoor activities and exercise) while it is also expected, especially in Westwood, that subjects tend to exercise along paths closer to their home thereby explaining the negative coefficient for the weighted distance to home.}

\begin{table}[t]
    \centering
    {\fontsize{8}{8}\selectfont
    \begin{tabular}{lccccccc}
    \toprule
        \multirow{2}{*}{\textbf{Param.}} & \multicolumn{2}{c}{\textbf{Model (\ref{modSpec1}) without temporal process}} & \multicolumn{2}{c}{\textbf{Model (\ref{modSpec1}) with temporal process}}\\
         &  Point & Interval &  Point & Interval\\
        \midrule
        Intercept               & -2.750 & (-2.754, -2.746)   & -2.92 & (-2.94, -2.91)      \\
        Eth. White              & -0.128 & (-0.146, -0.111)    & -0.190    & (-0.258,  -0.125)    \\
        Eth. Other    & 0.122 & (0.110, 0.134)    & 0.128    & (0.077,  0.178)    \\
        Eth. Latin-American     & 0.259 & (0.247, 0.271)   & 0.314    & (0.264,  0.362)    \\
        Eth. Black/African/Caribbean    & 0.263 & (0.248, 0.278)    & 0.400    & (0.340,  0.461)    \\
        Sex Male                & -0.348 & (-0.358, -0.338)    & -0.298    & (-0.338, -0.258)    \\
        Normal weight                     & 0.121 & (0.110, 0.132)    & 0.297    & (0.252, 0.343)     \\
        Over weight                     & 0.351 & (0.330, 0.372)    & 0.482    & (0.398, 0.566)     \\
        Obese                     & 0.220 & (0.181, 0.258)    & 0.401    & (0.241, 0.560)     \\
        Age (25-34]             & -0.387 & (-0.398, -0.377) & -0.320   & (-0.362, -0.279)   \\
        Age (34-45]             & 0.080 & (0.064, 0.097) & 0.125   & (0.057, 0.191)   \\
        Age (45-70]             & -0.105 & (-0.132, -0.079) & -0.091   & (-0.192, 0.006)   \\
        Dist. from home        & -0.135 & (-0.142, -0.128) & -0.074   & (-0.102, -0.046)   \\
        Slope       & 0.052 & (0.047, 0.0.56) & -0.003   & (-0.12, 0.005)   \\
        Dist. to parks        & -0.221 & (-0.227, -0.214) & -0.066   & (-0.089, -0.043)   \\
        NDVI       & 0.226 & (0.221, 0.231) & 0.010   & (0.004, 0.015)   \\
        $\sigma^2$              &         &                    & 2.266    & (2.237, 2.297)     \\
        $\phi$                  &         &                    & 0.718    & (0.704, 0.731)     \\
        $\tau^2$                &   2.10      &     (2.08, 2.13)               & 0.050    & (0.048, 0.053)     \\
         \midrule
         \textbf{Metric} & Out-of-sample & In-sample & Out-of-sample & In-sample & \\
         \midrule 
        DIC &  \multicolumn{2}{c}{17'588'058} &    \multicolumn{2}{c}{973'329} & \\

         Coverage   & 0.95         & 0.95         & 0.93         & 0.99        \\
         RMSPE (r)  & 1.44 (0.68)  & 1.44 (0.68) & 0.55 (0.09) & 0.1 (0.003)  \\
         PIW        & 5.69         & 5.69         & 2         & 1.18        \\
         \bottomrule    
    \end{tabular}
    }
    \caption{Parameter estimates and model performance metrics for model \eqref{modSpec1} with and without the temporal process.}
    \label{tab:ResTableSpApp}
\end{table}

The estimate %practical range is $\hat{r}_S=1/1.092=0.916$ for the S-Spline, and 
of %\textcolor{purple}
{the temporal decay parameter} $\phi$ implies that the %\textcolor{purple}
{temporal correlation drops to $0.05$ in about $3/\hat{\phi} \approx 4.3$ minutes, where $\hat{\phi} \approx 0.7$ is the posterior median of $\phi$}. Unsurprisingly, including the spatial effect and the temporal process improves predictive performances (RMSPE or PIW in Table~\ref{tab:ResTableSpApp}) over a model including only spatial effects (excluding the temporal process). The model incorporating the temporal process delivers satisfactory coverage and outperforms its competitor in all of the other indices for the training and testing data. 

\begin{figure}[htb]
\centering
\begin{subfigure}[b]{.5\linewidth}
     \centering
    \includegraphics[width=\textwidth]{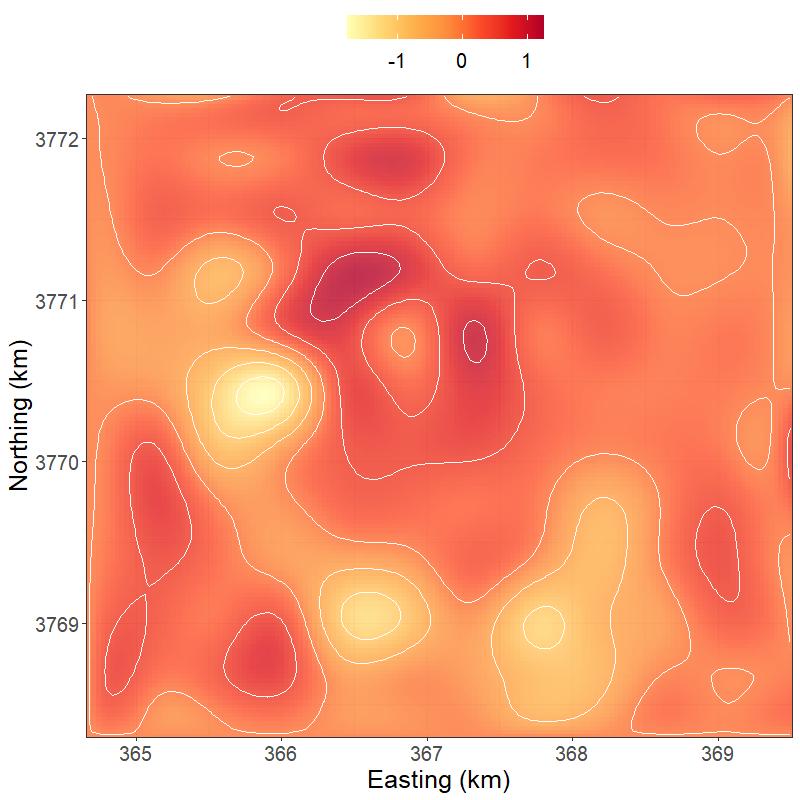}
    \caption{}
    \label{fig:EstSSonWW}
\end{subfigure}%
\begin{subfigure}[b]{.5\linewidth}
     \centering
    \includegraphics[width=\textwidth]{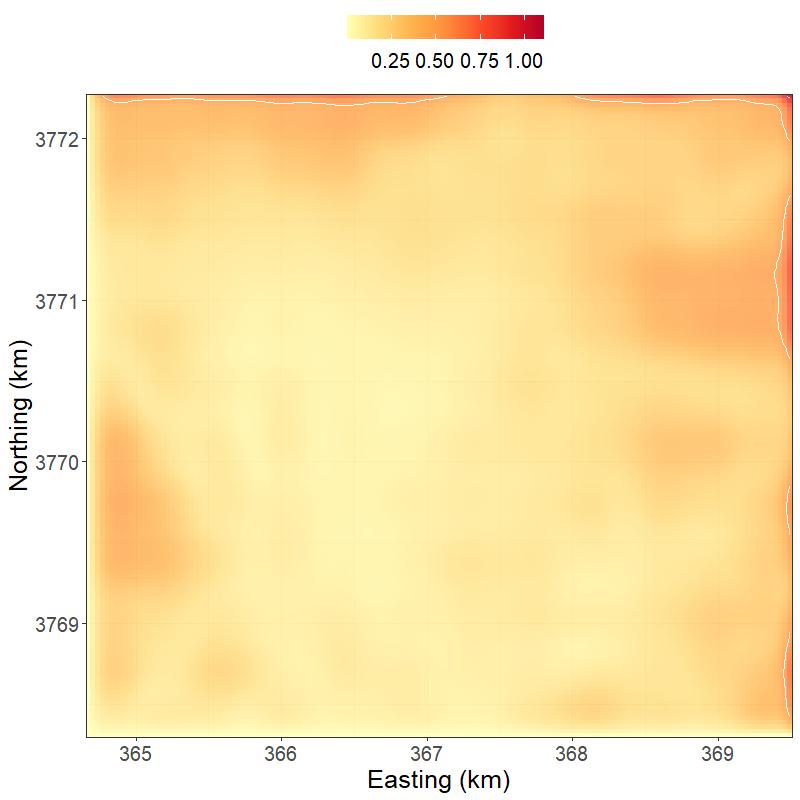}
    \caption{}
    \label{fig:WidthSSonWW}
\end{subfigure} 
\caption{(a) Spatially smoothed estimates from a shrinkage spline over Westwood, Los Angeles; (b) Standard deviation for the shrinkage spline.}
\label{fig:WWAreaSplines}
\end{figure}

Figure~\ref{fig:EstSSonWW} shows the estimated spatial surface, while Figure~\ref{fig:WidthSSonWW} presents the width of the posterior predictive intervals. The map clearly evinces zones (darker shades of red highlighted with white contours) that tend to depict high levels of physical activity. For example, the largest dark red blob in the north center-left almost perfectly tracks the UCLA campus boundary reflecting a campus environment with active mobility (walking, running, biking). Other zones of high activity identify with locations where more participants in the study live, including those residing in student dorms (northwest corner) and residential areas immediately around and in the predefined Westwood/UCLA study area (such as the south central zone) or Century City shopping center (to the east). Lighter shades (orange) correspond to areas that are less developed (open space), such as the areas in the north east; or they are areas with a high degree of transportation infrastructure and traffic (e.g., toward the western boundary). These correspond to highways (such as the Interstate-405 highway or other vehicular transportation corridors) that often have lower levels of activity because they inhibit outdoor physical activities due to noise, pollution, safety, etc. %To sum up, areas in red are known areas of high physical activity (e.g., where vehicle traffic is restricted---such as campus in the top center; century city mall in the bottom right), high residential areas, or green spaces, recreational areas and parks. Blue areas are undeveloped spaces, or areas along transportation corridors (such as highways and on- and off-ramps). 
Our analysis reveals three additional high activity areas that are not gleaned from non-spatial models: the Los Angeles National Veteran Park; the Century City shopping center and the Stone Canyon Park. The color gradient closely follows the spatial characteristics of the Westwood neighborhood and reveal how spatial patterns can impact physical activity behavior after accounting for variation attributable to known explanatory variables.
%Results highlight a central area, which coincides with the UCLA campus,  with large physical activity level. We also see a contour area (mostly composed of the main roads around the campus) with lower activity levels. 

\begin{figure}[ht]
    \centering
    \begin{subfigure}[b]{.95\textwidth}
    \includegraphics[width = .9\textwidth]{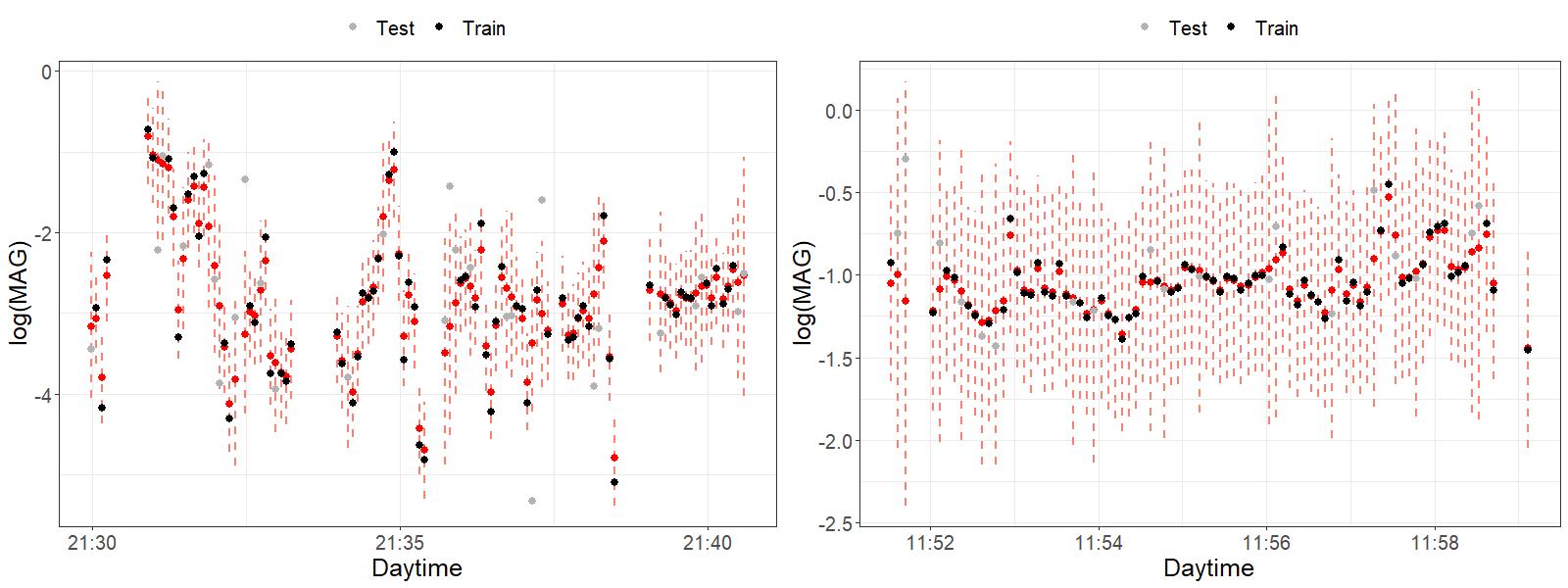}
    \caption{}
    \label{esttraj_time}
    \end{subfigure}
     \begin{subfigure}[b]{.95\textwidth}
    \includegraphics[width = .9\textwidth]{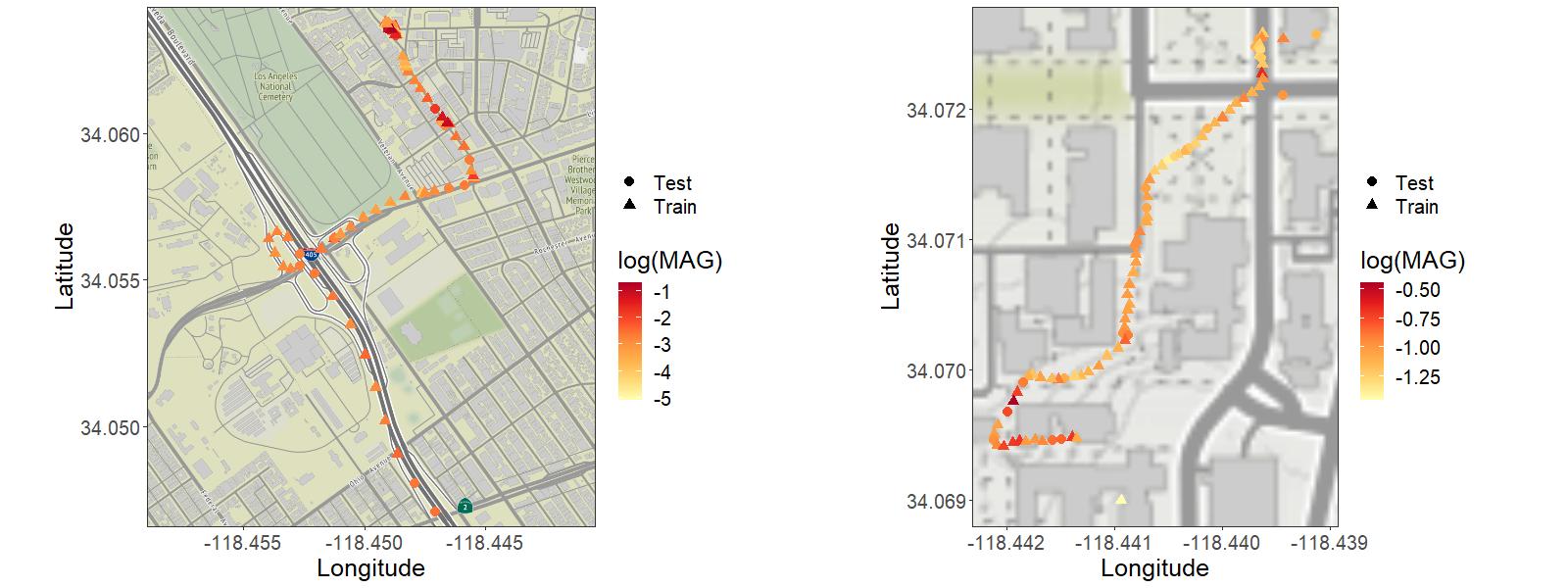}
    \caption{}
    \label{esttraj_space}
    \end{subfigure}
    \caption{Estimated log(MAG) for two randomly selected individuals: (a) estimated log(MAG) (red points) and 95\% prediction intervals (red dashed line) for each point within the observed time-windows; (b) including the location.}
    \label{fig:esttraj}
\end{figure}

Figure~\ref{fig:esttraj} shows two examples of observed (left) and reconstructed (right) MAGs along trajectories carved out by two subjects. We find a good degree of agreement between the two plots, and the ability of our model to recover the $\log$(MAG) in locations where it has not been observed. The reliability of the predictions can be demonstrated through different metrics and, unsurprisingly, %\textcolor{purple}
{accommodating} spatial effects and the temporal process improves predictive performances as measured by MSPE or PIW. %Compared to the linear regression neglecting the temporal dependence, it provides satisfactory coverage and largely outperforms it in all the other indices for the training and testing dataset. 
We deliver these personalized trajectory plots for every subject in the study and also predict personalized MAGs for each subject along any new trajectory. This enables personalized recommendations based upon an individual's health attributes including suggestions for more effective paths to follow for optimal physical activities, while also informing community level interventions in the built environment.

% \begin{figure}[t]
% \centering
% \begin{subfigure}[b]{\linewidth}
%   \centering
%   \includegraphics[trim= 0 100 0 100, clip, width=\textwidth]{images/204_17363_MAGFixedScale_pathtogether.pdf}
%   \caption{}
%   %\label{fig:obslocWW}
%   \end{subfigure}
%   \begin{subfigure}[b]{\linewidth}
%   \centering
%   \includegraphics[trim=0 100 0 100, clip, width=\textwidth]{images/566_17637_MAGFixedScale_pathtogether.pdf}
%   \caption{}
%   %\label{fig:obslocWW}
%   \end{subfigure}
% \caption{Two randomly chosen $lMAG$ trajectories over Westwood from individual 204 (a) and individual 566 (b). Observed trajectories with gaps are seen on the left panels; spatially reconstructed (predicted) trajectories are seen on the right panels.}
% \label{fig:mapsNew}
% \end{figure}

\section{Discussion}\label{secDisc}
We have devised a Bayesian modeling framework to conduct fully model-based inference for high-resolution accelerometer data over trajectories compiled from the PASTA-LA study. %This framework allows us to analyze and predict PA levels for an individual over trajectories of daily mobility. 
%The application motivates a somewhat unconventional spatiotemporal analysis. We depart from the more customary spatiotemporal process formulation and motivate dependence through temporal processes, while accounting for spatial variation using splines. 
Our key data analytic developments included (i) modeling dependence over trajectories; (ii) accounting for subject-specific spatial-temporal variation for daily mobility; and (iii) predicting or interpolating PA levels across trajectories; and (iv) identify zones of high physical activity in Westwood, Los Angeles. Our spatiotemporal analysis offers richer inference and reveals relationships between physical activity levels and a variety of factors, both at the subject level (e.g., personal attributes) and as a function of space and time. %We achieve scalability by using a DAG-based temporal process and introduced spatial effects through a tensor product basis of spline functions. 
The temporal process was able to effectively extract the features of the data at finer resolutions, while the spatial splines accounted for residual spatial heterogeneity. Accommodating both temporal dependence and spatial heterogeneity demonstrably improved predictive ability and enabled us to effectively delineate zones of high physical activity. 
%This fully model based methodology supports proper identification of features (individual-specific or environmental) predictive of physical activity along trajectories while accounting for uncertainty. 
Furthermore, the ability of the model to pool information across individuals at all time points allows us to infer about those who present sparsely observed space-time points (due to technical issues or protocol violation). In particular, we can interpolate and infer about PA levels with full uncertainty quantification and ensure the desired coverage by our prediction intervals. %\textcolor{purple}
{The methods we develop can be adapted to model animal tracking and be compared to existing spatial models \citep[see, e.g.,][]{hedley2004transect}.}

Recent public health reviews call for interdisciplinary technological advances to more effectively measure spatiotemporal energetics of activity spaces in obesity and chronic disease research \citep{james2016, kestens2017, drewnowski2020}. Individual-level data, at aggregate, can be used to identify anchor points for physical activity and reveal causal pathways between built environment exposures and health. Our work is a novel contribution demonstrating methodologies to answer these pressing research questions.

Our analysis also resolves practical difficulties in using actigraph data. It is not cost-effective to deploy research-grade GlobalSat GPS and Actigraph units as they are very expensive and continued usage requires heavy staff involvement. Our methods can be applied to analogous, but less complete, data derived from smart phones and smart watches, then such devices could be deployed in much larger studies with much larger sample sizes at a fraction of the cost. Given the spatiotemporal nature of outdoor PA research, our ability to predict in areas of data missingness drastically improve inference related to the impacts of the built and natural environments on physical activity and active mobility.

%While our approach offers trajectory-based inference for actigraph data, 
We recognize that there are several avenues for further research. Substantive investigations pertaining to the PASTA-LA study will focus on %\textcolor{purple}
{the impact of intervention schemes designed to promote physical activities} and ask questions related to controlling for weather while estimating the impact of the intervention. Our DAG-based approach for scalable temporal processes can be further enriched with recent developments
\citep{katzfuss2021, peruzzi2022highly}, although any of the methods reviewed and evaluated by \cite{heaton2019case} can be incorporated into our framework. 
%\textcolor{purple}
{We also recognize a wealth of future research surrounding wearable devices and actigraphy data. Examples include methodological advancements in clustering of trajectories according to different levels of physical activity and creating personalized health recommendation systems for patients with regard to trajectories (e.g., walking, running or biking routes) that will be most appropriate for them. Related to the clustering of trajectories, one can also pursue model-based learning about individual effects from the extent of (appropriately quantifies) spatial overlap in trajectories and discerning them from spatial effects  Finally, there is possible merit in modeling both the non-idle and idle times with a more comprehensive hierarchical specification \citep{bai2018two}. The joint modeling could be achieved using Mixture Models, Hidden Markov Models, or the modeling of multivariate Gaussian censored outcomes \citep{de2005bayesian, molstad2021gaussian}. Combining such approaches with efficient estimation strategies is the major challenge, which will be tackled in future developments of this work.}
% Finally, there is possible merit in modeling the activity counts in each axis jointly and relaxing the assumptions of Gaussianity using recent developments in multivariate spatiotemporal count models and for non-Gaussian outcomes \citep[see, e.g.][]{bradley2018computationally, bradley2020conjugate}.

%%%%%%%%%%%%%%%%%%%%%%%%%%%%%%%%%%%%%%%%%%%%%%
%% Support information (funding), if any,   %%
%% should be provided in the                %%
%% Acknowledgements section.                %%
%%%%%%%%%%%%%%%%%%%%%%%%%%%%%%%%%%%%%%%%%%%%%%
% \section*{Supplementary Materials}
% The online supplement (supplementary material) provides additional simulation experiments expanding upon the analysis presented here and further details on MCMC specifications. Computer programs executing the models and reproducing the analysis are also provided.

\section*{Acknowledgments}
The authors thank the Editor, Associate Editor and two anonymous reviewers for several helpful comments and suggestions. Sudipto Banerjee was supported, in part, by National Science Foundation (NSF) under grants  DMS-2113778, DMS-1916349 and IIS-1562303. Sudipto Banerjee and Michael B. Jerrett have been supported by the National Institute of Environmental Health Sciences (NIEHS) under grants R01ES030210 and 5R01ES027027. The authors also acknowledge support from the NIOSH Education Research Center, the Center for Occupational and Environmental Health, and the UCLA Department of Transportation for this work. Finally, the authors acknowledge support for the survey administration from the EU Physical Activity through Sustainable Transport Approaches (PASTA) team members \url{https://www.pastaproject.eu/}

\section*{Funding}
The work of the authors have been supported in part by National Science Foundation (NSF) under grants NSF/DMS 1916349 and NSF/IIS 1562303, and by the National Institute of Environmental Health Sciences (NIEHS) under grants R01ES030210 and 5R01ES027027.

%	\subsection*{Acknowledgements}

%\vspace{2cm}
	%\begin{center}
	%    {\Huge Appendix}
	%\end{center}
	
	%\appendix
	
	%\input{appendix_core}
	%\spacingset{1.25}
 \bibliographystyle{plainnat}

% Appendix

\section*{Supplementary Material}
\subsection*{Algorithms, Simulations and Data Analysis}

Supplementary Material is available in \cite{alaimo2023aoasSupplement}. It includes further information about data processing, relationships between vector magnitude of acceleration and metabolic equivalent of task, technical details on some of the algorithms using the temporal NNGP, additional simulation experiments and some further analysis of the actigraph data.

\subsection*{Computer programs}

Computer programs developed for implementing the models in the paper for the R statistical computing environment are available as Supplementary Material in the form of a compressed folder EfficientTNNGPforActigraph-main.zip. This can also be downloaded from a GitHub repository {\small \url{https://github.com/minmar94/EfficientTNNGPforActigraph}}

\end{document}